\documentclass[twocolumn,showpacs,amsmath,amssymb,prd]{revtex4}
\usepackage{graphicx}
\newcommand{\beq}{\begin{equation}}
\newcommand{\eeq}{\end{equation}}
\newcommand{\beqn}{\begin{eqnarray}}
\newcommand{\eeqn}{\end{eqnarray}}

\newcommand{\ve}[1]{\mbox{\boldmath $#1$}}

\newcommand{\half} {{1\over 2}}
\newcommand{\bsrho}{$(b^2/\rho_0)_{\rm horizon}$ }

\usepackage{graphicx}% Include figure files
\usepackage{dcolumn}% Align table columns on decimal point
\usepackage{bm}% bold math
\usepackage{epsf}

\begin{document}
\title{Relativistic magnetohydrodynamics in dynamical spacetimes: \\
A new AMR implementation}

\author{Zachariah B.\ Etienne}

\author{Yuk Tung Liu}

\author{Stuart L.\ Shapiro}
\altaffiliation{Also at Department of Astronomy \& NCSA, University of Illinois
at Urbana-Champaign, Urbana, IL 61801}

\affiliation{Department of Physics, University of Illinois at
Urbana-Champaign, Urbana, IL~61801}

\begin{abstract}
We have written and tested a new general relativistic
magnetohydrodynamics (GRMHD) code, capable of evolving MHD fluids in
dynamical spacetimes with adaptive-mesh refinement (AMR).  Our code  
solves the Einstein-Maxwell-MHD system of coupled equations in full 
3+1 dimensions, evolving the metric via the Baumgarte-Shapiro
Shibata-Nakamura (BSSN) formalism and the MHD and magnetic
induction equations via a conservative, high-resolution shock-capturing 
scheme.  The
induction equations are recast as an evolution equation for the
magnetic vector potential, which exists on a grid that is staggered
with respect to the hydrodynamic and metric variables. 
The divergenceless constraint $\ve{\nabla}\cdot \ve{B}=0$ is 
enforced by the curl of the vector potential.  Our MHD scheme 
is fully compatible with AMR, so that fluids at AMR refinement
boundaries maintain $\ve{\nabla}\cdot \ve{B}=0$.  In simulations with
uniform grid spacing, our MHD scheme is numerically equivalent to a
commonly used, staggered-mesh constrained-transport scheme.  We present
code validation test results, both in Minkowski and curved spacetimes. 
They include magnetized shocks, nonlinear
Alfv\'en waves, cylindrical explosions, cylindrical rotating
disks, magnetized Bondi tests, and the collapse of a magnetized rotating
star. Some of the more stringent tests involve black holes. 
We find good agreement between analytic and numerical
solutions in these tests, and achieve convergence at the expected order.
\end{abstract}


\maketitle

\section{Introduction}
\label{intro}

Magnetized fluids in dynamical, strongly curved spacetimes play a
central role in many systems of current interest in relativistic
astrophysics.  Such fluids may generate gamma-rays in gamma-ray bursts (GRBs),
destroy differential rotation in nascent neutron
stars arising from stellar core collapse or binary neutron star
merger, form jets and influence disk dynamics around black
holes, affect magnetorotational collapse of massive stars, etc. 
Many of these systems are promising sources of gravitational radiation
for detection by laser interferometers such as LIGO, VIRGO, TAMA, GEO
and LISA.  Some also emit electromagnetic radiation, such as gamma-ray
bursts, magnetized disks around black holes in active galactic nuclei (AGNs) 
and quasars, and binary supermassive black holes coalescing in ambient 
magnetized plasma.  Accurate, self-consistent modeling
of these systems requires a computational scheme capable of
simultaneously accounting for magnetic fields, relativistic magnetohydrodynamics
(MHD) and relativistic gravitation.

Over the past several years, we have developed a robust numerical
scheme in 3+1 dimensions that evolves the Einstein equations of general relativity
for the gravitational field (metric), coupled to
the equations of relativistic MHD
for the matter and Maxwell's equations for a magnetic
field~\cite{dlss05}. Our approach is based on the BSSN
(Baumgarte-Shapiro-Shibata-Nakamura) formalism to evolve the
metric~\cite{SN,BS},
a high-resolution, shock-capturing (HRSC) scheme to handle
the fluids, and a constrained-transport scheme to treat
magnetic induction~\cite{t00}. This GRMHD code has been subjected to a
rigorous suite of numerical tests to check and calibrate its
validity~\cite{dlss05}.  The code has been applied to explore
a number of important dynamical scenarios in relativistic astrophysics, including the
collapse of magnetized, differentially-rotating
hypermassive neutron stars to black holes~\cite{dlsss06a,dlsss06b,ssl08},
the collapse of rotating stellar cores to neutron stars~\cite{slss06},
the collapse of rotating, supermassive stars and massive
Population~III stars to black holes~\cite{lss07}, magnetized binary neutron star 
merger~\cite{lset08}, 
binary black hole-neutron
stars~\cite{eflstb08,elsb09}, and the merger of 
binary black holes in gaseous environments~\cite{fls10}.
The purpose of this paper is to present a generalization of
our current GRMHD scheme that is compatible with adaptive mesh
refinement (AMR).

Many problems in relativistic astrophysics require 
numerical simulations covering a large range of lengthscales. 
For example, to follow the final merger of a compact binary system 
with a total mass $M$, a lengthscale of $\sim M/30$ needs to be
resolved to treat the strong-field, near-zone regions reliably.  On the other
hand, accurate gravitational wave calculations at lengthscale $\sim M$ must be
performed far in the weak-field wave-zone at radius $r \gtrsim 100M$.  AMR allows for
sufficient resolution to be supplied to areas of the computational domain
as needed, thus enabling us to resolve strong- and weak-field domains
efficiently.

One of the most subtle issues in evolving the MHD equations 
is the preservation of the divergenceless constraint 
($\ve{\nabla}\cdot \ve{B}=0$) during the 
evolution.  When evolving the induction equations, numerical
truncation error leads to violations of the divergenceless
constraint, resulting in unphysical plasma transport orthogonal to the
magnetic field,  as well as violations of energy and momentum
conservation (see e.g., \cite{bb80,bal_sp99,t00}).  
In simulations using a uniformly spaced grid, ``constrained-transport'' 
schemes (see e.g., \cite{eh88,t00}) are commonly used to maintain the
divergenceless constraint.  In these schemes, special finite-differencing
representations of the induction equations are implemented
to preserve a particular numerical representation of the divergence of the 
magnetic field to roundoff error.  
In simulations using AMR grids, both constrained-transport 
schemes and the hyperbolic divergence-cleaning scheme~\cite{dkkmsw02,ahln06}  
have been used.  In the hyperbolic divergence-cleaning scheme, 
a generalized Lagrange multiplier (GLM) scalar is coupled to the
system of MHD and induction equations.  No special finite-differencing
treatment is needed in solving the GLM-MHD system of equations. When
they appear, divergence errors of the magnetic field are both
propagated and damped away in the GLM-MHD scheme.

In the development of our AMR GRMHD code, we first tried 
the hyperbolic divergence-cleaning scheme, due to 
its straightforward implementation.  We found that this 
scheme works well in the absence of black holes. 
One of the most commonly-adopted methods for evolving black holes is
the moving puncture technique~\cite{rit06,goddard06a}, in which the
physical singularity in the  
black hole interior is avoided by the use of the puncture 
gauge conditions.  However, a coordinate singularity is present in 
the computational domain around which accurate numerical evolution 
is difficult to achieve. It has been demonstrated that the BSSN scheme, coupled 
with the puncture gauge conditions, guarantee that any inaccurate 
data in the black hole interior will not propagate out of the 
horizon~\cite{eflsb07,bsstdhp07,bdsst09}. We find that this 
property is preserved in the presence of 
hydrodynamic matter. However, it is no longer the case in the 
GLM-MHD scheme. In fact, we find that even in the 
Cowling approximation in which the metric is fixed, inaccurate 
data in the black hole interior can propagate out of the 
horizon in the GLM-MHD systems of equations. This problem 
may be overcome via black hole excision and  
applying appropriate ingoing boundary conditions at the excision 
boundary. (See~\cite{clr08} for a discussion of constraint preserving 
boundary conditions for Newtonian MHD.)
%%This is more easily done if the GLM-MHD equations 
%%are coupled to the Einstein equations written in the 
%%first-order hyperbolic form (see e.g. \cite{bona89,bona92,kst01,lskor06}), 
%%where the characteristics of the system of equations 
%%are relatively easy to calculate.

In developing an algorithm for maintaining $\ve{\nabla}\cdot \ve{B}=0$
that is compatible with the moving puncture technique, we focused on 
constrained-transport schemes.  That was the approach adopted in 
our earlier unigrid implementation~\cite{dlss05}.
A uniform-resolution,
constrained-transport scheme may be used on each individual AMR
refinement level.  However, maintaining the divergenceless constraint
at refinement level {\it boundaries} requires that special interpolations be
performed during prolongation/restriction.  Such
prolongation/restriction operators have been
devised~\cite{balsara01,balsara09}, but must be fine-tuned to the
particular AMR implementation.  
In this paper, we propose an alternative, AMR-compatible constrained-transport scheme. 
Our scheme is based on the constrained-transport scheme 
described in~\cite{zbl03}. In this scheme, the magnetic 
induction equation is recast as an evolution equation 
for the magnetic vector potential. The divergence-free 
magnetic field is computed via the curl of the vector 
potential. The evolution of the vector potential is carried 
out in the same HRSC framework as other hydrodynamic variables. 
This scheme is numerically equivalent to 
the commonly used constrained-transport schemes 
based on a staggered mesh algorithm~\cite{eh88}. This 
scheme is readily generalized to an AMR grid. Unlike the 
magnetic field, the vector potential is not constrained, 
and so any interpolation scheme can be used during 
prolongation and restriction, thus enabling its use with {\it any} AMR 
algorithm. 

We have performed several tests on our new AMR constrained-transport 
scheme. We find that it works well even in 
black-hole spacetimes. Inaccurate data generated in the black hole 
interior stay inside the horizon. Hence our scheme is 
compatible with the moving puncture technique. 

The structure of this paper is as follows. In Sec.~\ref{sec:formal}, 
we describe our formalism, focusing on the derivation of the 
evolution equation for the magnetic vector potential. 
Then we describe our numerical scheme to evolve the coupled 
Einstein-Maxwell-MHD equations (Sec.~\ref{sec:numerical}).
Next we present several stringent code tests, including one- and two-dimensional
shocks, magnetized Bondi accretion and the collapse of a magnetized
rotating star (Sec.~\ref{sec:tests}). Finally, we summarize our work
in Sec.~\ref{sec:summary} and discuss applications of our new code to
study various interesting problems in relativistic astrophysics.

\section{Formalism}
\label{sec:formal}

The formulation and numerical approach adopted in this paper are basically
the same as those already reported in our previous work~\cite{dlss05,eflstb08,elsb09}, to which
the reader may refer for details. Here we introduce our notation,
summarize our method, and focus on the derivation of the
evolution equation for the magnetic vector potential in the ideal
MHD limit, which is the basis of our new AMR constrained-transport scheme.
Geometrized units ($G = c = 1$) are adopted throughout. 
Greek indices denote all four
spacetime dimensions (0, 1, 2, and 3), and Latin indices imply spatial
parts only (1, 2, and 3).

\subsection{Metric evolution and gauge conditions}
We use the standard 3+1 formulation of general relativity and decompose
the metric into the following form:
\beq
  ds^2 = -\alpha^2 dt^2
+ \gamma_{ij} (dx^i + \beta^i dt) (dx^j + \beta^j dt) \ .
\eeq
The fundamental variables for the metric evolution are the spatial
three-metric $\gamma_{ij}$ and extrinsic curvature $K_{ij}$. We adopt
the BSSN formalism~\cite{SN,BS} in which
the evolution variables are the conformal exponent $\phi
\equiv \ln (\gamma)/12$, the conformal 3-metric $\tilde
\gamma_{ij}=e^{-4\phi}\gamma_{ij}$, three auxiliary functions
$\tilde{\Gamma}^i \equiv -\tilde \gamma^{ij}{}_{,j}$, the trace of
the extrinsic curvature $K=\gamma_{ij} K^{ij}$,
 and the trace-free part of the conformal 
extrinsic curvature $\tilde A_{ij} \equiv e^{-4\phi}(K_{ij}-\gamma_{ij} K/3)$.
Here, $\gamma={\rm det}(\gamma_{ij})$ is the determinant of 
the spatial metric. The full spacetime metric $g_{\mu \nu}$
is related to the three-metric $\gamma_{\mu \nu}$ by $\gamma_{\mu \nu}
= g_{\mu \nu} + n_{\mu} n_{\nu}$, where the future-directed, timelike
unit vector $n^{\mu}$ normal to the time slice can be written in terms
of the lapse $\alpha$ and shift $\beta^i$ as $n^{\mu} = \alpha^{-1}
(1,-\beta^i)$. The evolution equations of these BSSN variables are
given by Eqs.~(9)--(13) in~\cite{eflstb08}. 
It has been suggested that
evolving $\chi=e^{-4\phi}$ or $W=e^{-2\phi}$ instead of $\phi$
gives more accurate results in binary black hole simulations (see
e.g.~\cite{bghhst08,FAU_BBH,rit08}). Our code is capable 
of evolving these variables. Kreiss-Oliger dissipation is sometimes 
added in the BSSN evolution equations to reduce 
high-frequency numerical noise associated with AMR refinement
interfaces~\cite{goddard06}. It is also found that Kreiss-Oliger dissipation 
is sometimes useful in hydrodynamic simulations involving a black hole in 
a dynamical spacetime~\cite{bgr08,elsb09}. 

We adopt standard puncture gauge conditions to evolve the lapse 
and shift: an advective
``1+log'' slicing condition for the lapse and a
``Gamma-freezing'' condition for the shift~\cite{GodGauge}. 
The evolution equations for these quantities are given by 
Eqs.~(2)--(4) in~\cite{elsb09}.

\subsection{Evolution of electromagnetic fields}

The electromagnetic stress-energy tensor $T^{\mu\nu}_{\rm em}$ is given
by
\beq
 T^{\mu\nu}_{\rm em} = \frac{1}{4\pi} \left(
F^{\mu\lambda}F^{\nu}{}_{\lambda} - \frac{1}{4}
g^{\mu\nu}F_{\alpha\beta}F^{\alpha\beta} \right) \ .
\eeq
We decompose the Faraday tensor $F^{\mu\nu}$ as
\beq
  F^{\mu \nu} = n^{\mu} E^{\nu} - n^{\nu} E^{\mu} + n_{\gamma}
\epsilon^{\gamma \mu \nu \delta} B_{\delta} \ ,
\label{eq:FabEB}
\eeq
so that $E^{\mu}$ and $B^{\mu}$ are the electric and magnetic
fields measured by an observer normal to the spatial slice $n^{\mu}$. 
Both fields
are purely spatial ($E^{\mu}n_{\mu} = B^{\mu}n_{\mu} = 0$), and
one can easily show that
\beqn
  E^{\mu} = F^{\mu \nu} n_{\nu} \ \ \ , \ \ \
  B^{\mu} = \frac{1}{2} \epsilon^{\mu \nu \kappa \lambda} n_{\nu}
F_{\lambda \kappa} = n_{\nu} F^{* \nu \mu} \ ,
\label{eq:EB}
\eeqn
where
\beq
   F^{* \mu \nu} = \frac{1}{2} \epsilon^{\mu \nu \kappa \lambda}
F_{\kappa \lambda}
\eeq
is the dual of $F^{\mu \nu}$.

Along with the electromagnetic field, we also assume the presence of
a perfect fluid with rest-mass density $\rho_0$, pressure $P$, and 4-velocity
$u^{\mu}$, so that the total stress-energy tensor is
\beq
  T^{\mu \nu} = \rho_0 h u^{\mu} u^{\nu} + P g^{\mu \nu}
  + T_{\rm em}^{\mu \nu}\ ,
\eeq
where the specific enthalpy $h$ is related to the specific internal energy
$\epsilon$ by $h = 1 + \epsilon + P/\rho_0$. 
The electric and magnetic fields measured by an observer comoving
with the fluid are [cf.\ Eq.~(\ref{eq:EB})]
\beqn
  E^{\mu}_{(u)} = F^{\mu \nu} u_{\nu}  \ \ \ , \ \ \
  B^{\mu}_{(u)} = u_{\nu} F^{* \nu \mu} \ .
\label{eq:EuBu}
\eeqn

For many applications of interest in relativistic astrophysics, one can assume
perfect conductivity. In this ideal MHD limit, Ohm's law yields
the MHD condition:
\beq
u_{\mu}F^{\mu\nu} = 0\ , \label{cond:MHD}
\eeq
which is equivalent to the statement that the electric field observed 
in the fluid's rest frame vanishes ($E^{\mu}_{(u)}=0$). 
In this limit, the total stress-energy tensor is given by 
\beq
  T^{\mu \nu} = (\rho_0 h +b^2) u^{\mu} u^{\nu} + \left( P + \frac{b^2}{2}
\right) g^{\mu \nu} - b^{\mu} b^{\nu} \ ,
\eeq
where $b^\mu = B^{\mu}_{(u)}/\sqrt{4\pi}$ and $b^2=b^\mu b_\mu$. The vector 
$b^\mu$ is related to $B^\mu$ by (see~\cite{dlss05} for a derivation) 
\beq
  b^\mu = -\frac{P^{\mu}{}_{\nu} B^{\nu}}{n_{\nu} u^{\nu}\sqrt{4\pi}} \ ,
\label{eq:bmu}
\eeq
where $P_{\mu \nu} = g_{\mu \nu} + u_\mu u_\nu$ is a projection tensor.

The evolution equation for the magnetic field in a perfectly
conducting MHD fluid can be obtained in
conservative form by taking the dual of Maxwell's equation
$F_{[\mu \nu,\lambda]}=0$.  One finds
\beq
  \nabla_{\nu} F^{* \mu \nu} = \frac{1}{\sqrt{-g}} \partial_{\nu}
(\sqrt{-g}\, F^{*\mu \nu}) = 0 \ , \label{Maxwell}
\eeq
where $\sqrt{-g} = \alpha \sqrt{\gamma}$. The time component of 
Eq.~(\ref{Maxwell}) gives the no-monopole constraint 
\beq
\partial_j \tilde{B}^j =0 \ , \label{eq:nomonopole}
\eeq
where 
\beq
  \tilde{B}^i = \sqrt{\gamma} B^i \ .
\label{eq:Btildei}
\eeq
The spatial components of Eq.~(\ref{Maxwell}) give the
magnetic induction equation, which can be written as
\beq
  \partial_t \tilde{B}^i + \partial_j (v^j \tilde{B}^i - v^i
\tilde{B}^j)=0 \ ,
\label{eqn:induction}
\eeq
where $v^i = u^i/u^0$. 

The induction equation can be recast as
\beq
  \partial_t \tilde{B}^i = \tilde{\epsilon}^{ijk} \tilde{\epsilon}_{klm} 
\partial_j ( v^l \tilde{B}^m) \ ,
\label{eqn:induction2}
\eeq
where both $\tilde{\epsilon}^{ijk}$ and $\tilde{\epsilon}_{ijk}$ 
denote the permutation symbol, i.e. they are equal to 1 if $ijk$ 
are in even permutation of (1,2,3), $-1$ if in odd 
permutation, and 0 if any two of the indices are equal. 
The divergenceless constraint~(\ref{eq:nomonopole}) implies that 
$\tilde{B}^i$ can be derived from a vector potential $A_i$: 
\beq
  \tilde{B}^i = \tilde{\epsilon}^{ijk} \partial_j A_k \ .
\label{eq:Ai}
\eeq
It follows
from Eqs.~(\ref{eq:Btildei}) and (\ref{eq:Ai}) that
\beq
  B^i = \epsilon^{ijk} \partial_j A_k \ ,
\label{eq:BfromA}
\eeq
where $\epsilon^{ijk} = \tilde{\epsilon}^{ijk}/\sqrt{\gamma} =
n_\mu \epsilon^{\mu ijk}$ is the three-dimensional Levi-Civita
tensor associated with $\gamma_{ij}$. Equation~(\ref{eq:BfromA})
can be derived in a more general framework, as shown in~\cite{bs03}.

The induction equation~(\ref{eqn:induction2}) will be satisfied 
automatically if $A_i$ satisfies the evolution equation 
\beq
  \partial_t A_i = \tilde{\epsilon}_{ijk} v^j \tilde{B}^k \ .
\label{eq:indAi}
\eeq
It is clear that the evolution equations for $A_i$ are not unique, since
there are gauge degrees of freedom in the electromagnetic 4-vector 
potential.  The general evolution equation for $A_i$ in the ideal 
MHD limit is obtained by combining Eqs.~(33) and (46) in~\cite{bs03}:
\beq
  \partial_t A_i = \tilde{\epsilon}_{ijk} v^j \tilde{B}^k 
 - \partial_i (\alpha \Phi - \beta^j A_j) \ ,
\eeq
where $\Phi$ is the electromagnetic scalar potential. Hence the 
evolution equation~(\ref{eq:indAi}) is equivalent to choosing 
the electromagnetic gauge condition
\beq
  \Phi = \frac{1}{\alpha} (C + \beta^j A_j) \ ,
\label{eq:scalarPhi}
\eeq
where $C$ is a constant. In Minkowski spacetime, in which $\alpha=1$ and 
$\beta^i=0$, the gauge condition reduces to $\Phi=C$. 

The scalar potential is only needed if 
one wishes to compute the electric field $E^i$. However, in the 
ideal MHD limit, the condition $u_\mu F^{\mu \nu}=0$ relates 
$E^i$ to $B^i$ and $v^i$: 
$\alpha E_i = -\epsilon_{ijk}(v^j+\beta^j)B^k$. Therefore, it is 
not necessary to keep track of the scalar potential $\Phi$ in the 
ideal MHD limit.

In the nonrelativistic limit, Eq.~(\ref{eqn:induction2}) reduces to 
\beq
  \partial_t \ve{B} = \ve{\nabla} \times (\ve{v} \times \ve{B}) \ 
\eeq
and Eqs.~(\ref{eq:BfromA}) and (\ref{eq:indAi}) reduce to 
\beq
  \ve{B} = \ve{\nabla} \times \ve{A} \ \ \ \ \ , \ \ \ \ \ 
  \partial_t \ve{A} = \ve{v} \times \ve{B} \ .
\eeq
The ideal MHD condition becomes $\ve{E} = -\ve{v}\times \ve{B}$.

In our new AMR constrained-transport scheme, the induction 
equation is evolved via Eq.~(\ref{eq:indAi}).
The divergence-free magnetic field is then computed using 
Eq.~(\ref{eq:BfromA}). The numerical implementation 
will be described in Sec.~\ref{sec:numerical}. 

\subsection{Evolution of the hydrodynamics fields}

The stress-energy tensor for a magnetized plasma in the ideal MHD 
limit is 
\beq
 T^{\mu \nu} = (\rho_0 h +b^2) u^{\mu} u^{\nu} + \left( P + \frac{b^2}{2}
\right) g^{\mu \nu} - b^{\mu} b^{\nu} \ . 
\eeq
Our evolution variables are 
\beqn
 &&\rho_* \equiv - \sqrt{\gamma}\, \rho_0 n_{\mu} u^{\mu},
\label{eq:rhos} \\
&& \tilde{S}_i \equiv -  \sqrt{\gamma}\, T_{\mu \nu}n^{\mu} \gamma^{\nu}_{~i}, \\
&& \tilde{\tau} \equiv  \sqrt{\gamma}\, T_{\mu \nu}n^{\mu} n^{\nu} - \rho_*.
\label{eq:S0} 
\eeqn
The evolution equations are derived from the rest-mass conservation 
law $\nabla_\mu (\rho_0 u^\mu)=0$ and conservation of energy-momentum 
$\nabla_\mu T^{\mu \nu}=0$. These result in the 
continuity, momentum and energy equations~\cite{dlss05} 
\beqn
  \partial_t \rho_* + \partial_j (\rho_* v^j) & = & 0 \ , 
\label{eq:continuity} \\
\partial_t \tilde{S}_i
 + \partial_j (\alpha \sqrt{\gamma}\, T^j{}_i) & = & \frac{1}{2} \alpha \sqrt{\gamma}
\, T^{\alpha \beta} g_{\alpha \beta,i} \ , \\
  \partial_t \tilde{\tau} + \partial_i ( \alpha^2 \sqrt{\gamma}\, T^{0i}
-\rho_* v^i) & = & s \ , \label{eq:energy} 
\eeqn
where the source term in the energy equation is given by 
\beqn
s &=& -\alpha \sqrt{\gamma}\, T^{\mu \nu} \nabla_{\nu} n_{\mu}  \cr
   &=& \alpha \sqrt{\gamma}\, [ (T^{00}\beta^i \beta^j + 2 T^{0i} \beta^j
+ T^{ij}) K_{ij} \cr
 & & - (T^{00} \beta^i + T^{0i}) \partial_i \alpha ]\ .
\eeqn
To complete the system of equations, the fluid equation of state (EOS)
is specified.  Our code currently implements a hybrid EOS of the
form~\cite{jzm93}
\beq
  P(\rho_0,\epsilon) = P_{\rm cold}(\rho_0) + (\Gamma_{\rm th}-1)
\rho_0 [\epsilon-\epsilon_{\rm cold}(\rho_0)] \ ,
\eeq
where $P_{\rm cold}$ and $\epsilon_{\rm cold}$ denote the cold component 
of $P$ and $\epsilon$ respectively, and $\Gamma_{\rm th}$ is a constant 
parameter which determines the conversion efficiency of kinetic 
to thermal energy at shocks.  The function $\epsilon_{\rm
  cold}(\rho_0)$ is related to $P_{\rm cold}(\rho_0)$ by the first law
of thermodynamics,
\beq
  \epsilon_{\rm cold}(\rho_0) = \int \frac{P_{\rm cold}(\rho_0)}{\rho_0^2} d\rho_0 \ .
\eeq
In the code tests presented in this paper, 
we adopt the $\Gamma$-law EOS $P=(\Gamma-1)\rho_0 \epsilon$. This 
corresponds to setting $P_{\rm cold}=\kappa\rho_0^\Gamma$ (with constant 
$\kappa$) and $\Gamma_{\rm th}=\Gamma$.

\section{Numerical Implementation}
\label{sec:numerical}

\begin{table}
\caption{Storage location on grid of the magnetic field $B^i$ and vector potential $A_i$}
\begin{tabular}{cc}
\hline
  Variable & storage location \\
\hline
  $B^x$, $\tilde{B}^x$ & $(i+\half,j,k)$ \\
  $B^y$, $\tilde{B}^y$ & $(i,j+\half,k)$ \\ 
  $B^z$, $\tilde{B}^z$ & $(i,j,k+\half)$ \\
  $A_x$ & $(i,j+\half,k+\half)$ \\
  $A_y$ & $(i+\half,j,k+\half)$ \\
  $A_z$ & $(i+\half,j+\half,k)$ \\
\hline
\end{tabular}
\label{tab:staggeredBA}
\end{table}

We adopt Cartesian coordinates in our 3+1 simulations. Equatorial
symmetry (i.e. symmetry with respect to the reflection $z\rightarrow
-z$) is imposed when appropriate to save computational time.  All
the BSSN and hydrodynamical 
variables are stored at grid points $(i,j,k)$. Magnetic field $B^i$ and 
vector potential $A_i$ are stored at staggered grid points as summarized 
in Table~\ref{tab:staggeredBA}. 

The BSSN equations are evolved using a finite-differencing scheme. Our code 
currently supports second, fourth, and sixth order spatial finite-differencing. 
In a spacetime containing black holes, we typically use a fourth or sixth 
order finite-differencing scheme. Our code is embedded in
the Cactus parallelization framework~\cite{Cactus}, with 
time-stepping managed by the {\tt MoL} (Method of Lines) thorn, which 
supports various explicit time-stepping algorithms.  Typically, we use
the fourth-order Runge-Kutta method in time when evolving spacetimes
containing black holes.

We use the Carpet~\cite{Carpet} infrastructure to implement moving-box
adaptive mesh refinement. In all AMR simulations presented here, 
second-order temporal prolongation is employed, coupled with fifth-order
spatial prolongation for evolution variables stored on the unstaggered grid. 
The memory allocation for the staggered variables are the same as 
the unstaggered ones. The staggering is incorporated in 
our code in the evolution steps. Different spatial prolongation and 
restriction schemes have to be applied on the staggered evolution 
variables $A_i$ to account for the different relative positions 
of these variables on adjacent refinement levels. We currently use a third-order 
Lagrangian scheme for interpolating these variables, but it can be easily generalized 
to other higher-order schemes, as well as more sophisticated schemes 
such as the essentially non-oscillatory (ENO)~\cite{ho87} and weighted 
essentially non-oscillatory (WENO)~\cite{loc94,js96} schemes. We plan 
to investigate these alternative schemes in the future.

\subsection{MHD evolution}

The technique for evolving the BSSN equations is described in our earlier 
papers~\cite{eflstb08,elsb09,les09}, so we focus here on our MHD
evolution technique, which is based on an HRSC scheme. 
The goal of this part of the numerical evolution is to determine the
fundamental MHD variables ${\bf P} = (\rho_0,P,v^i,B^i)$,
called the {\it ``primitive'' variables}, at future times, given
initial values of ${\bf P}$. 
The evolution equations~(\ref{eqn:induction}), 
(\ref{eq:continuity})--(\ref{eq:energy}) are written in conservative
form: 
\beq
  \partial_t \ve{U} + \ve{\nabla}\cdot \ve{F} = \ve{S} \ ,
\label{eq:coneq}
\eeq
where $\ve{U}(\ve{P}) = (\rho_*,\tilde{\tau},\tilde{S}_i,\tilde{B}^i)$ are the 
``conserved'' variables, and the flux 
$\ve{F}(\ve{P})$ and source $\ve{S}(\ve{P})$ do not contain derivatives 
of the primitive variables, although they are explicit functions of the 
metric and its derivatives.

%Our evolution variables are $\rho_*$, $\tilde{\tau}$, $\tilde{S}_i$ and 
%$A_i$. Note that $B^i$, $\tilde{B}^i$ and $A_i$ are stored on a 
%staggered grid as indicated in Table~\ref{tab:staggeredBA}. The evolution 
%of $A_i$ is different from that of $\rho_*$, $\tilde{\tau}$, and $\tilde{S}_i$ 
%and will be described in the next subsection. 

Equation~(\ref{eq:coneq}) may be 
evolved using a finite-volume or finite-difference scheme. A finite-volume 
scheme evolves the volume-averaged variables, whereas a finite-difference 
scheme evolves the point-valued variables. Our adopted constrained-transport 
scheme is based on a finite-volume algorithm. In a second-order 
scheme, there is no distinction 
between these two types of methods since the volume average and 
the gridpoint value are the same to second order.  Since the metric is evolved 
using a finite-difference scheme, care must be taken to evolve 
the MHD and induction equations using a higher-order finite-volume scheme. 
One solution is to evolve the volume averaged conservative variables 
$\bar{\ve{U}}$ from the point-value primitive variables $\ve{P}$ using 
a finite-volume algorithm. Next the updated point-value $\ve{U}$ is 
computed from the updated  
volume average quantity $\bar{\ve{U}}$ to the desired order of 
accuracy. The updated point-value $\ve{P}$ is then computed from the 
updated point-value $\ve{U}$ and metric 
quantities through primitives inversion. In this paper, we only consider second-order 
schemes for simplicity.  Higher-order schemes are planned for the
future, and important extra steps necessary to go beyond 
second-order will be reviewed in this section.

Equation~(\ref{eq:coneq}) can be written in a finite-volume form by integrating 
it over a cell volume. We obtain 
\beqn
 \partial_t \bar{\ve{U}}_{i,j,k} + \frac{(\Delta_x \langle \ve{F}\rangle)_{i,j,k}}{\Delta x} 
 + \frac{(\Delta_y \langle \ve{F}\rangle)_{i,j,k}}{\Delta y} && \cr
 + \frac{(\Delta_z \langle \ve{F}\rangle)_{i,j,k}}{\Delta z} = \bar{\ve{S}}_{i,j,k} \ ,\ \  && 
\label{eq:Udot}
\eeqn
where 
\beq
  (\Delta_x \langle \ve{F}\rangle)_{i,j,k} \equiv 
\langle\ve{F}\rangle_{i+{1\over 2},j,k}-\langle\ve{F}\rangle_{i-{1\over 2},j,k}
\eeq
and similarly for operators $\Delta_y$ and $\Delta_z$. 
We note that only a subset of $\ve{U}$, i.e. $\rho_*$, $\tilde{\tau}$ 
and $\tilde{S}_i$, is evolved using Eq.~(\ref{eq:Udot}). The evolution  
of $\tilde{B}^i$ will be described in the next subsection. The bracket 
$\langle \rangle$ denotes a surface average. For example, 
\beq
  \langle \ve{F}\rangle_{i+{1\over 2},j,k} \equiv 
 \frac{1}{\Delta y \Delta z} 
\int_{y_j^-}^{y_j^+} dy
\int_{z_k^-}^{z_k^+} dz \,
\ve{F} \left( x_i^+,y,z \right) \ ,
\eeq
where $x_i^{\pm} = x_i \pm \Delta x/2$, $y_j^{\pm} = y_j \pm \Delta y/2$ and 
$z_k^{\pm} = z_k \pm \Delta z/2$. The fluxes $\langle \ve{F}\rangle_{i,j+{1\over 2},k}$ 
and $\langle \ve{F}\rangle_{i,j,k+{1\over 2}}$ are defined in the same way except that 
the surfaces to be averaged are in the $x$-$z$ plane and $x$-$y$ plane, respectively. 
The surface averaged flux $\langle \ve{F} \rangle$ and the point-value flux $\ve{F}$ 
are the same to second-order accuracy. To implement a higher-order scheme, 
one needs to compute not only the point-value $\ve{F}$ at the zone interface to the desired order, 
but also $\langle \ve{F} \rangle$ from the point-value $\ve{F}$ to 
the desired order of accuracy.

The computation of the fluxes is 
basically the same as described in~\cite{dlss05}. It involves 
the reconstruction step and the Riemann solver step. In the 
reconstruction step, primitive variables at the zone interface 
are reconstructed. A slope-limited interpolation scheme from 
the zone center gives $\ve{P}_R$ and $\ve{P}_L$, the primitive 
variables at the right and left side of each zone interface, respectively. 
We usually employ the piecewise parabolic method (PPM)~\cite{PPM}  
or the monotonized central (MC)~\cite{MC} reconstruction scheme, 
but in some problems involving strong discontinuities 
a more diffusive scheme such as the minmod reconstruction scheme 
must be used (see Sec.~\ref{sec:2d-min-tests}).
Since $B^i$ is staggered (as shown in Table~\ref{tab:staggeredBA}), 
each $B^i$ at one of the zone interfaces need not be computed. 
From $\ve{P}_R$ and $\ve{P}_L$, we compute the fluxes 
$\ve{F}_R$ and $\ve{F}_L$, the ``conservative'' 
variables $\ve{U}_R$ and $\ve{U}_L$, as well as two pairs of 
characteristic velocities $c_{\pm}^R$ and $c_{\pm}^L$ at each
zone interface (see Sec.~IIIB of~\cite{dlss05} for details). 

The next step is the Riemann solver step. We employ the HLL 
(Harten, Lax, and van Leer) approximate Riemann
solver~\cite{HLL} in which the HLL flux is given by 
\beq
  F^{\rm HLL} = {c^- F_R + c^+ F_L
  - c^+ c^- (u_R - u_L) \over
  c^+ + c^- }\ ,
\label{eq:Fhll}
\eeq
where $c^{\pm} = \max(0,\pm c_{\pm}^R,\pm c_{\pm}^L)$. Our code 
also has the option of using the single-speed, local Lax-Friedrichs (LLF),  
or central-upwind, flux,
\beq
  F^{\rm LLF} = \frac{1}{2} [ F_R + F_L - c (u_R - u_L) ] \ ,
\label{eq:Fllf}
\eeq
where $c=\max(c^+,c^-)$. 

The accuracy of the resulting flux depends on the reconstruction 
scheme and Riemann solver. In a smooth flow, MC reconstruction 
results in a second-order accurate point-value flux $\ve{F}$, whereas PPM is third-order. 
However, these two schemes reduce to first-order in a discontinuous 
flow (e.g. shocks) or at local extrema of $\ve{P}$. 
As mentioned above, even in a smooth flow where PPM gives third-order 
accurate point-value $\ve{F}$, $\langle \ve{F} \rangle$ has to be 
computed from the point-value $\ve{F}$ to third order to 
achieve an overall third-order accuracy. 

\subsection{Constrained transport scheme}
\label{sec:amr-ct}
In this subsection, the standard constrained-transport 
scheme based on the staggered algorithm~\cite{eh88} is reviewed briefly. 
Next we introduce the vector potential 
method described in~\cite{zbl03}. These two approaches give
numerically identical results for schemes in which the time
integration and spatial derivatives commute.

The evolution variables for the magnetic field in the standard 
constrained-transport scheme is the surface averaged field $\langle \tilde{B}^i \rangle$ 
defined in the same way as the surface averaged fluxes:  
\beqn
  \langle \tilde{B}^x \rangle_{i+{1 \over 2},j,k} &\equiv& \frac{1}{\Delta y \Delta z} 
\int_{y_j^-}^{y_j^+} dy
\int_{z_k^-}^{z_k^+} dz \,
\tilde{B}^x \left( x_i^+,y,z \right) \ \ \label{eq:Bxtavg}  \\
  \langle \tilde{B}^y \rangle_{i,j+{1\over 2},k} &\equiv& \frac{1}{\Delta x \Delta z} 
\int_{x_i^-}^{x_i^+} dx
\int_{z_k^-}^{z_k^+} dz \,
\tilde{B}^y \left( x,y_j^+,z \right) \ \ \label{eq:Bytavg}  \\
\langle \tilde{B}^z \rangle_{i,j,k+{1\over 2}} &\equiv& \frac{1}{\Delta x \Delta y} 
\int_{x_i^-}^{x_i^+} dx
\int_{y_j^-}^{j_j^+} dy \,
\tilde{B}^z \left( x,y,z_k^+ \right), \ \ \label{eq:Bztavg}
\eeqn
Integrating the magnetic constraint equation $\partial_j \tilde{B}^j=0$ 
over a cell volume gives the finite-volume equation for the constraint 
\beq
  \frac{(\Delta_x \langle \tilde{B}^x \rangle)_{i,j,k}}{\Delta x} 
 + \frac{(\Delta_y \langle \tilde{B}^y \rangle)_{i,j,k}}{\Delta y}
 + \frac{(\Delta_z \langle \tilde{B}^z \rangle)_{i,j,k}}{\Delta z} = 0\ .
\label{eq:divBfv}
\eeq
To derive the finite-volume equation for the magnetic induction equation, 
we first rewrite Eq.~(\ref{eqn:induction2}) as 
\beqn
  \partial_t \tilde{B}^x &=& -\partial_y {\cal E}^z + \partial_z {\cal E}^y \ ,
\label{eq:Bxtdot} \\ 
  \partial_t \tilde{B}^y &=& -\partial_z {\cal E}^x + \partial_x {\cal E}^z \ ,
\label{eq:Bytdot}\\
  \partial_t \tilde{B}^z &=& -\partial_x {\cal E}^y + \partial_y {\cal E}^x \ , 
\label{eq:Bztdot}
\eeqn
where 
\beqn
  {\cal E}^x &=& -v^y \tilde{B}^z + v^z \tilde{B}^y \ , \\
  {\cal E}^y &=& -v^z \tilde{B}^x + v^x \tilde{B}^z \ , \\
  {\cal E}^z &=& -v^x \tilde{B}^y + v^y \tilde{B}^x \ .
\eeqn
We next define the line averaged ${\cal E}^i$ as 
\beqn
  \hat{\cal E}^x_{i,j+\half,k+\half} &\equiv & \frac{1}{\Delta x} 
\int_{x_i^-}^{x_i^+} {\cal E}^x(x,y_j^+,z_k^+) dx \ , \\
  \hat{\cal E}^y_{i+\half,j,k+\half} &\equiv & \frac{1}{\Delta y}
\int_{y_j^-}^{y_j^+} {\cal E}^y(x_i^+,y,z_k^+) dy \ , \\
  \hat{\cal E}^z_{i+\half,j+\half,k} &\equiv & \frac{1}{\Delta z}
\int_{z_k^-}^{z_k^+} {\cal E}^z(x_i^+,y_j^+,z) dz \ .
\eeqn
Note that $\hat{\cal E}^i$ is staggered in the same way as $A_i$ 
(see Table~\ref{tab:staggeredBA}). The finite-volume equations 
for the magnetic induction are obtained by integrating 
Eq.~(\ref{eq:Bxtdot}) over the cell surface normal to the $x$-direction, 
integrating Eq.~(\ref{eq:Bytdot}) over the cell surface normal to the $y$-direction, 
and integrating Eq.~(\ref{eq:Bztdot}) over the cell surface normal to the $z$-direction: 
\beqn
  \partial_t \langle \tilde{B}^x \rangle_{i+\half,j,k} &=& 
\frac{(\Delta_z \hat{\cal E}^y)_{i+\half,j,k}}{\Delta z}
-\frac{(\Delta_y \hat{\cal E}^z)_{i+\half,j,k}}{\Delta y} \ , \ \ \
\label{eq:indBtx} \\
  \partial_t \langle \tilde{B}^y \rangle_{i,j+\half,k} &=&
\frac{(\Delta_x \hat{\cal E}^z)_{i,j+\half,k}}{\Delta x}
-\frac{(\Delta_z \hat{\cal E}^x)_{i,j+\half,k}}{\Delta z} \ , \ \ \
\label{eq:indBty} \\
  \partial_t \langle \tilde{B}^z \rangle_{i,j,k+\half} &=&
\frac{(\Delta_y \hat{\cal E}^x)_{i,j,k+\half}}{\Delta y}
-\frac{(\Delta_x \hat{\cal E}^y)_{i,j,k+\half}}{\Delta x} \ . \ \ \
\label{eq:indBtz}
\eeqn
It is straightforward to verify that Eqs.~(\ref{eq:indBtx})--(\ref{eq:indBtz}) 
imply that the time derivative of the left hand side of Eq.~(\ref{eq:divBfv}) 
vanishes. Hence a finite-volume 
scheme that evolves Eqs.~(\ref{eq:indBtx})--(\ref{eq:indBtz}) preserves  
the constraint~(\ref{eq:divBfv}) to roundoff error, provided that the 
initial data $\langle \tilde{B}^i \rangle$ satisfy the constraint. 

To evolve Eqs.~(\ref{eq:indBtx})--(\ref{eq:indBtz}), 
$\ve{\cal E}$ has to be computed at the zone edge. 
The computation is similar to that of the flux $\ve{F}$ described 
in the previous subsection. Since $B^i$ is staggered (as specified in 
Table~\ref{tab:staggeredBA}), computation of each ${\cal E}^i$ at the zone 
edge requires reconstruction of $B^i$ along one direction. However, 
since $v^i$ is stored at the zone center, two independent one-dimensional 
reconstructions are necessary, as pointed out in~\cite{zbl03}. 
The HLL and Lax-Friedrichs formulas for ${\cal E}^z$ at the zone edge 
are given by~\cite{zbl03} 
\beqn
  ({\cal E}^z)^{\rm HLL} = \frac{c^+_x c^+_y {\cal E}^z_{LL}
+ c_x^+ c_y^- {\cal E}^z_{LR} + c_x^- c_y^- {\cal E}^z_{RL}
+ c_x^- c_y^- {\cal E}^z_{RR}}{(c_x^+ + c_x^-)(c_y^+ + c_y^-)} && \cr 
 + \frac{c_x^+ c_x^-}{c_x^+ + c_x^-}
( \tilde{B}^y_R-\tilde{B}^y_L) 
- \frac{c_y^+ c_y^-}{c_y^+ + c_y^-}
( \tilde{B}^x_R-\tilde{B}^x_L)\ \ \  && 
\label{eq:Ezhll}
\eeqn
and
\beqn
  ({\cal E}^z)^{\rm LLF} &=& \frac{1}{4}({\cal E}^z_{LL} + {\cal E}^z_{LR} 
+ {\cal E}^z_{RL} + {\cal E}^z_{RR}) \cr 
&& + \frac{c_x}{2}(\tilde{B}^y_R-\tilde{B}^y_L) - \frac{c_y}{2}
(\tilde{B}^x_R-\tilde{B}^x_L) \ , 
\label{eq:Ezllf}
\eeqn
which are the generalizations of Eqs.~(\ref{eq:Fhll}) and (\ref{eq:Fllf}). 
In the above formulas, ${\cal E}^z_{LR}$ denotes 
the reconstructed left state in the $x$-direction and right state in the 
$y$-direction. Other symbols involving ${\cal E}^z$ are interpreted 
in the similar fashion. $\tilde{B}^y_R$ and $\tilde{B}^y_L$ denote 
the reconstructed right and left state of $\tilde{B}^y$ in the $x$-direction; 
$\tilde{B}^x_R$ and $\tilde{B}^x_L$ denote the reconstructed right and left 
state in the $y$-direction. The $c^{\pm}_x$ and $c^{\pm}_y$ should be 
computed by taking the maximum characteristic speed among the four 
reconstructed states. However, we set them equal to the maximum 
over the two neighboring interface values for simplicity, as suggested 
in~\cite{zbl03}. In the LLF formula, $c_x$ and $c_y$ are set to the 
maximum of $c_x^{\pm}$ and $c_y^{\pm}$, respectively. The formula 
for $({\cal E}^x)^{\rm HLL}$ is obtained from Eq.~(\ref{eq:Ezhll}) by 
permuting the indices 
$z \rightarrow x$, $x \rightarrow y$ and $y\rightarrow z$, 
whereas the formula for $({\cal E}^y)^{\rm HLL}$ is obtained from 
Eq.~(\ref{eq:Ezhll}) by permuting the indices 
$z \rightarrow y$, $x \rightarrow z$ and $y \rightarrow x$. 
The same rule applies for $({\cal E}^x)^{\rm LLF}$ and 
$({\cal E}^y)^{\rm LLF}$. 
The reconstructed point-value ${\cal E}^i$ at the zone edge is 
the same as the line averaged value $\hat{\cal E}^i$ to second-order. 
If one wishes to go beyond second-order, $\hat{\cal E}^i$ has to be 
computed from ${\cal E}^i$ to the desired order of accuracy. 

We now describe the vector potential method proposed in~\cite{zbl03}, which 
has been adopted for our AMR constrained-transport scheme.  We first define 
the line averaged vector potential $\hat{A}_i$ exactly the same way 
as $\hat{\cal E}^i$: 
\beqn
  (\hat{A}_x)_{i,j+\half,k+\half} &\equiv & \frac{1}{\Delta x}
\int_{x_i^-}^{x_i^+} A_x(x,y_j^+,z_k^+) dx \ ,  \\
  (\hat{A_y})_{i+\half,j,k+\half} &\equiv & \frac{1}{\Delta y}
\int_{y_j^-}^{y_j^+} A_y(x_i^+,y,z_k^+) dy \ , \\
  (\hat{A_z})_{i+\half,j+\half,k} &\equiv & \frac{1}{\Delta z}
\int_{z_k^-}^{z_k^+} A_z(x_i^+,y_j^+,z) dz \ . 
\eeqn
It follows from Eq.~(\ref{eq:Ai}) that 
\beqn
  \langle \tilde{B}^x \rangle_{i+\half,j,k} &=& 
  \frac{(\Delta_y \hat{A}_z)_{i+\half,j,k}}{\Delta y} 
- \frac{(\Delta_z \hat{A}_y)_{i+\half,j,k}}{\Delta z} \ , \ \  
\label{eq:BxtfromA}\\ 
  \langle \tilde{B}^y \rangle_{i,j+\half,k} &=&
  \frac{(\Delta_z \hat{A}_x)_{i,j+\half,k}}{\Delta z}
- \frac{(\Delta_x \hat{A}_z)_{i,j+\half,k}}{\Delta x} \ , \ \  
\label{eq:BytfromA} \\
  \langle \tilde{B}^z \rangle_{i,j,k+\half} &=&
  \frac{(\Delta_x \hat{A}_y)_{i,j,k+\half}}{\Delta x}
- \frac{(\Delta_y \hat{A}_x)_{i,j,k+\half}}{\Delta y} \ . \ \ 
\label{eq:BztfromA}
\eeqn
It is easy to verify that the data $\langle \tilde{B}^i \rangle$ 
generated from $\hat{A}_i$ from the above formulas satisfy the 
constraint~(\ref{eq:divBfv}). In the vector potential method, 
the evolution variable is $\hat{A}_i$. The evolution equation 
is derived from Eq.~(\ref{eq:indAi}) and is given by 
\beqn
  \partial_t (\hat{A}_x)_{i,j+\half,k+\half} &=& 
-\hat{\cal E}^x_{i,j+\half,k+\half} \ , \label{eq:Axdot} \\
  \partial_t (\hat{A}_y)_{i+\half,j,k+\half} &=&
-\hat{\cal E}^y_{i+\half,j,k+\half} \ , \label{eq:Aydot} \\
  \partial_t (\hat{A}_z)_{i+\half,j+\half,k} &=&
-\hat{\cal E}^z_{i+\half,j+\half,k} \ . \label{eq:Azdot}
\eeqn
The value of ${\cal E}^i$ at the zone edge is computed in 
exactly the same way as the standard constrained-transport scheme, 
i.e., by using Eq.~(\ref{eq:Ezhll}) or Eq.~(\ref{eq:Ezllf}) 
for ${\cal E}^z$ and similar formulas for ${\cal E}^x$ and 
${\cal E}^y$. Having evolved $\hat{A}_i$, the updated conservative 
variables $\langle \tilde{B}^i \rangle$ are computed using 
Eqs.~(\ref{eq:BxtfromA})--(\ref{eq:BztfromA}). The divergence of 
$\langle \tilde{B}^i \rangle$ is therefore automatically guaranteed to
be zero to roundoff error. 

It is apparent that the vector potential method and the standard 
constrained-transport scheme are closely related. They both apply  
the same procedure of computing ${\cal E}^i$ at the zone edge. They 
both involve taking spatial derivatives (more precisely, the 
discretized curl operator) via the differencing 
operators $\Delta_x$, $\Delta_y$ and $\Delta_z$. The only difference 
between these two methods is that in the standard constrained-transport 
scheme, spatial derivatives are applied before time integration, whereas 
in the vector potential method spatial derivatives are applied after 
time integration. Since we employ the MoL algorithm in which 
spatial derivatives and time integration commute, the two methods 
give numerically identical results in simulations using 
a uniformly-spaced grid. We prefer to use the vector 
potential method in AMR simulations since $A_i$ is not constrained and so 
does not require special interpolation schemes during prolongation 
and restriction. 

During the MHD evolution steps,
values of $B^i$ at the zone center are also needed, which are currently
computed by simply taking the average of $B^i$ on the staggered grid. 
Taking the limit $\Delta x^i \rightarrow 0$ in Eq.~(\ref{eq:divBfv}), 
we see that $\langle \tilde{B}^i \rangle$ is always continuous in 
the $x^i$ direction (e.g.,\ even in the presence of shocks). 
Thus the averaging scheme generally gives a second-order 
accurate $B^i$ at the zone center. 
Higher-order schemes will require more sophisticated interpolation
algorithms.

\subsection{Recovery of primitive variables}

Having computed ${\bf U}$ at the new timestep, we need to recover 
${\bf P}$, the primitive variables on the new time level. 
This is not trivial because, although the relations ${\bf U}({\bf P})$
are analytic, the inverse relations ${\bf P}({\bf U})$ are not. 
For a $\Gamma$-law EOS $P=(\Gamma-1)\rho_0 \epsilon$, the inversion 
can be reduced to an eighth-order polynomial equation~\cite{zbl03,ngmz06}. 
In the absence of magnetic field, the equation can be further reduced to 
a quartic equation where an analytic solution is available. However, for a 
general EOS, the inversion must be solved numerically. Various inversion 
algorithms are studied extensively in~\cite{ngmz06}, and it has been found that 
the most efficient inversion technique is to solve two coupled nonlinear equations 
using the Newton-Raphson scheme. 

Our code supports three inversion schemes: the optimal 2D scheme described in~\cite{ngmz06}, 
a slightly modified analytic quartic solver from the GNU Scientific
Library (used for a $\Gamma$-law EOS in the absence of magnetic field), and our 
older scheme that solves four coupled nonlinear algebraic equations.

\subsection{Black hole interior}

We use the moving puncture technique to evolve spacetimes containing black holes.
The black hole spacetime singularity is avoided by the puncture gauge conditions, 
but a coordinate 
singularity (i.e.\ puncture) remains in the interior of each black hole on the 
computational domain. One nice property of the moving puncture method is that, 
although accurate evolution near the puncture is not maintained, inaccurate data 
do not propagate out of the black hole horizon. This method proves to be robust 
in the evolution of binary black holes and is widely used in the numerical 
relativity community. The moving puncture method has also been used in simulations 
involving hydrodynamic matter and MHD (see e.g., 
\cite{lset08,bgr08,yst08,skyt09,elsb09,fls10,bhbls10}).

One difficulty in handling MHD in the black hole interior is the loss 
of accuracy near the puncture. 
This can drive the ``conservative'' variables 
$\ve{U}$ out of physical range, resulting in unphysical primitive 
variables after inversion (e.g.\ negative pressure or even 
complex solutions). In the absence of magnetic fields, this can be 
avoided by enforcing the constraints~\cite{eflstb08}
\beqn
 |\tilde{S}|^2 \equiv \gamma^{ij} \tilde{S}_i \tilde{S}_j &<& \tilde{\tau} 
(\tilde{\tau}+2\rho_*) \ , {\rm and} \label{eq:S2con} \\ 
  \tilde{\tau} &>& 0 \ . \label{eq:taucon}
\eeqn
When the second condition is not met, we reset $\tau$ to a small 
positive number. When the first condition is violated we rescale 
$\tilde{S}_i$ so that its
new magnitude is $|\tilde{S}|^2=f
\tilde{\tau}(\tilde{\tau}+2\rho_*)$, where $f\leq 1$ is a parameter. 
This technique does not apply in the presence of magnetic fields. 
We instead apply a fix, first
suggested by Font {\it et al}~\cite{fmst00}, which consists of 
replacing the energy equation~(\ref{eq:energy}) by the cold EOS, 
$P = P_{\rm cold}(\rho_0)$ when solving the system of equations. 
This substitution guarantees a positive pressure.
In rare cases, this revised system also fails to give a solution
and we repair the zone by averaging from nearby zones  
(averaging is not applied to the magnetic field). 

When matter and magnetic fields fall into the black hole, the energy density 
near the puncture can be very high. This results in a large energy source term 
in the BSSN equation, which can cause the conformal related metric 
$\tilde{\gamma}_{ij}$ to lose positive definiteness near the puncture. 
This behavior eventually causes the code to crash. Hence, 
other techniques are sometimes used to stabilize the evolution in the 
black hole interior. For example, adding a Kreiss-Oliger dissipation 
in the black hole interior is found to be useful~\cite{bgr08,elsb09}, 
as well as setting an upper and lower limit on the pressure. In some 
MHD simulations, we find that setting the magnetic field to zero 
deep inside the horizon can stabilize the evolution 
(see Sec.~\ref{sec:bondi}).

\subsection{Low-density regions}
\label{sec:low-density}

If a pure vacuum were to exist anywhere in our computational domain, the MHD
approximation would not apply in this region, and the vacuum Maxwell
equations would need to be solved there. In many astrophysical scenarios, 
however, a sufficiently dense, ionized plasma will exist outside
the stars or disks, where MHD will remain valid in its force-free limit. 
As in many hydrodynamic and MHD simulations, we add a tenuous 
``atmosphere'' to cover the computational grid outside the star or disk. 
We maintain a density and pressure floor $\rho_{\rm atm}$ and $P_{\rm atm}$ 
in the atmosphere. We usually set $\rho_{\rm atm}=10^{-10} \rho_{\rm max}(0)$ 
and $P_{\rm atm} = P_{\rm cold}(\rho_{\rm atm})$, where $\rho_{\rm max}(0)$ 
is the maximum rest-mass density in the initial data. Throughout the
evolution, we impose limits on the atmospheric pressure to prevent
spurious heating and negative values of the internal energy when the 
density $\rho_0$ is smaller than a threshold $\rho_{\rm th}$. 
Specifically, we require $P_{\rm min}(\rho_0)\leq P \leq P_{\rm max}(\rho_0)$, 
where $P_{\rm max}(\rho_0)=10 P_{\rm cold}(\rho_0)$ and 
$P_{\rm min}(\rho_0)=P_{\rm cold}(\rho_0)/2$ when $\rho_0 <\rho_{\rm th}$. The 
value of $\rho_{\rm th}$ is usually set between $10 \rho_{\rm atm}$ 
and $100 \rho_{\rm atm}$. Setting $\rho_{\rm th}$ too high could cause 
unphysical effects in a simulation, such as spurious angular 
momentum loss~\cite{elsb09}. 

\subsection{Boundary conditions}

In simulations in which the spacetime is asymptotically flat, 
we apply Sommerfeld outgoing wave boundary conditions to the BSSN 
and gauge variables ${\bf f}$, i.e.,
\begin{equation}
\label{wavelike}
{\bf f}(r,t) = {r - \Delta r\over r}{\bf f}(r-\Delta r, t-\Delta T)
\end{equation}
at the outer boundary of our numerical grid.  Here $\Delta T$ is the
timestep and $\Delta r = \alpha e^{-2\phi}\Delta T$. 
In simulations in which hydrodynamic matter and plasma are initially localized, 
outflow boundary conditions are imposed on the hydrodynamic 
variables $\rho_0$, $v^i$ and $P$ (i.e., the variables are copied 
along the grid directions with the
condition that the velocities be positive or zero in the outer grid zones). 
For the magnetic field, we compute $A_i$ at the outer boundaries 
by either linear or quadratic extrapolation. The linear extrapolation 
is equivalent to copying $\tilde{B}^i$ to the outer boundary, whereas 
the quadratic extrapolation corresponds to linearly extrapolating 
$\tilde{B}^i$ to the outer boundary. 

Other boundary conditions are used in the code tests presented in 
this paper, which will be specified in each case. 

\section{Code tests}
\label{sec:tests}

\subsection{Minkowski spacetime MHD tests}

\subsubsection{One-dimensional tests} 

\begin{table*}
\caption{Initial states for 1D MHD tests.${}^{\rm a}$}
\begin{tabular}{c  c  c  c}
\hline \hline
 Test & Left state & Right State & $t_{\rm final}$ \\
\hline \hline
  Fast Shock & $u^i=(25.0,0.0,0.0)$ & $u^i=(1.091,0.3923,0.00)$ & 2.5 \\
 & $B^i/\sqrt{4\pi}=(20.0,25.02,0.0)$ &
 $B^i/\sqrt{4\pi}=(20.0,49.0,0.0)$ &  \\
  & $P=1.0$, $\rho_0=1.0$ & $P=367.5$, $\rho_0=25.48$ & \\
\hline
  Slow Shock & $u^i=(1.53,0.0,0.0)$ & $u^i=(0.9571,-0.6822,0.00)$ & 2.0 \\
 & $B^i/\sqrt{4\pi}=(10.0,18.28,0.0)$ &
 $B^i/\sqrt{4\pi}=(10.0,14.49,0.0)$ & \\
  & $P=10.0$, $\rho_0=1.0$ & $P=55.36$, $\rho_0=3.323$ & \\
\hline
  Switch-off Fast & $u^i=(-2.0,0.0,0.0)$ & $u^i=(-0.212,-0.590,0.0)$ & 1.0 \\
  Rarefaction & $B^i/\sqrt{4\pi}=(2.0,0.0,0.0)$ &
  $B^i/\sqrt{4\pi}=(2.0,4.71,0.0)$ & \\
  & $P=1.0$, $\rho_0=0.1$ & $P=10.0$, $\rho_0=0.562$ & \\
\hline
  Switch-on Slow & $u^i=(-0.765,-1.386,0.0)$ & $u^i=(0.0,0.0,0.0)$ & 2.0 \\
  Rarefaction & $B^i/\sqrt{4\pi}=(1.0,1.022,0.0)$ &
 $B^i/\sqrt{4\pi}=(1.0,0.0,0.0)$ & \\
 & $P=0.1$, $\rho_0=1.78\times 10^{-3}$ & $P=1.0$, $\rho_0=0.01$ & \\
\hline
  Shock Tube 1 & $u^i=(0.0,0.0,0.0)$ & $u^i=(0.0,0.0,0.0)$ & 1.0 \\
  & $B^i/\sqrt{4\pi}=(1.0,0.0,0.0)$ & $B^i/\sqrt{4\pi}=(1.0,0.0,0.0)$ & \\
  & $P=1000.0$, $\rho_0=1.0$ & $P=1.0$, $\rho_0=0.1$ & \\
\hline
  Shock Tube 2 & $u^i=(0.0,0.0,0.0)$ & $u^i=(0.0,0.0,0.0)$ & 1.0 \\
 & $B^i/\sqrt{4\pi}=(0.0,20.0,0.0)$ & $B^i/\sqrt{4\pi}=(0.0,0.0,0.0)$ & \\
 & $P=30.0$, $\rho_0=1.0$ & $P=1.0$, $\rho_0=0.1$ & \\
\hline
Nonlinear Alfv\'en wave${}^{\rm b}$ & $u^i=(0.0,0.0,0.0)$  &
$u^i=(3.70,5.76,0.00)$ & 2.0 \\
 & $B^i/\sqrt{4\pi}=(3.0,3.0,0.0)$ &
$B^i/\sqrt{4\pi}=(3.0,-6.857,0.0)$ & \\
 & $P=1.0$, $\rho_0=1.0$ & $P=1.0$, $\rho_0=1.0$ & \\
\hline \hline
\end{tabular}
\vskip 12pt
\begin{minipage}{12cm}
\raggedright
${}^{\rm a}$ {In all cases, the gas satisfies the $\Gamma$-law EOS with
$\Gamma = 4/3$. For the first 6 tests, the left state refers to $x<0$
and the right state, $x>0$.} \\
${}^{\rm b}$ {For the nonlinear Alfv\'en wave, the left and right
states are joined by a continuous function. See~\cite{k97}
or Appendix~B of~\cite{dlss05} for details.} \\
\end{minipage}
\label{tab:1Dtests}
\end{table*}

We perform a suite of one-dimensional MHD tests in Minkowski
spacetime, as described in~\cite{k99}. The initial configurations of
the tests are summarized in Table~\ref{tab:1Dtests}. We only perform
tests in which analytic solutions are available. In these 1D tests,
all variables are functions of $x$ only.  The divergenceless
constraint $\ve{\nabla}\cdot \ve{B}=0$ implies that $B^x$ is a
constant. For these tests, the initial magnetic field $\ve{B}$ can be derived
from the following vector potential:
\beqn
  A_x(x) &=& 0 \ , \\ 
  A_y(x) &=& \int_0^x B^z(x') dx' \ , \\ 
  A_z(x) &=& y B^x - \int_0^x B^y(x') dx' \ .
\eeqn

\begin{figure}
\includegraphics[width=9cm]{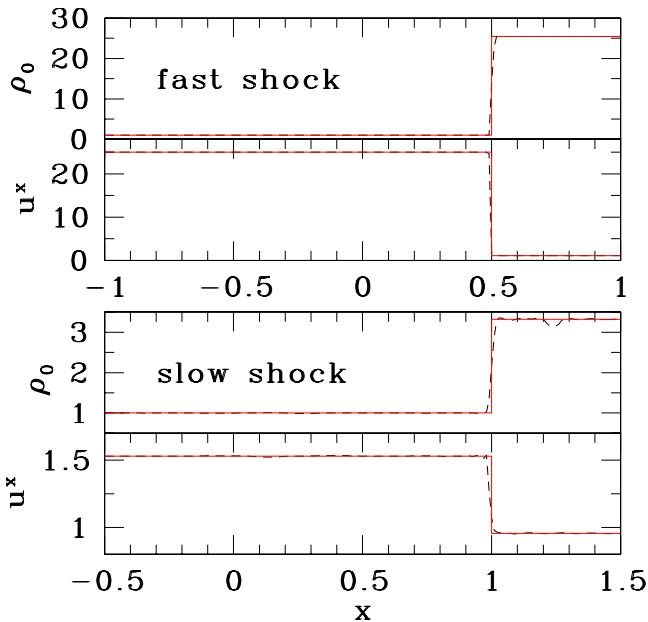}
\caption{1D fast and slow shock density and velocity profiles,
at $t=t_{\rm final}$ (see Table~\ref{tab:1Dtests}). 
Data from numerical simulations with resolution $\Delta x=0.01$ are
plotted with dashed (black) lines, and solid (red) lines 
denote the analytic solutions.}
\label{fig:fast_slow_shock}
\end{figure}

\begin{figure}
\includegraphics[width=9cm]{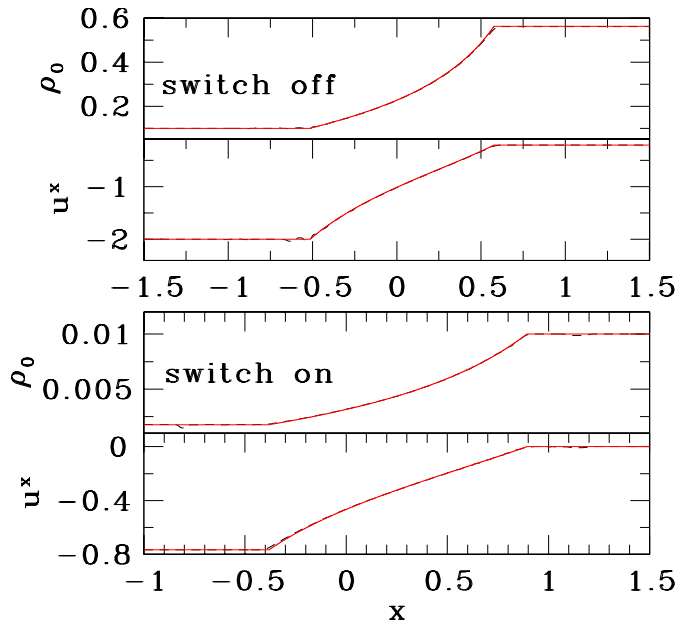}
\caption{Same as Fig.~\ref{fig:fast_slow_shock} but for the 1D switch-off 
and switch-on tests.}
\label{fig:switch_off_on}
\end{figure}

\begin{figure}
\includegraphics[width=9cm]{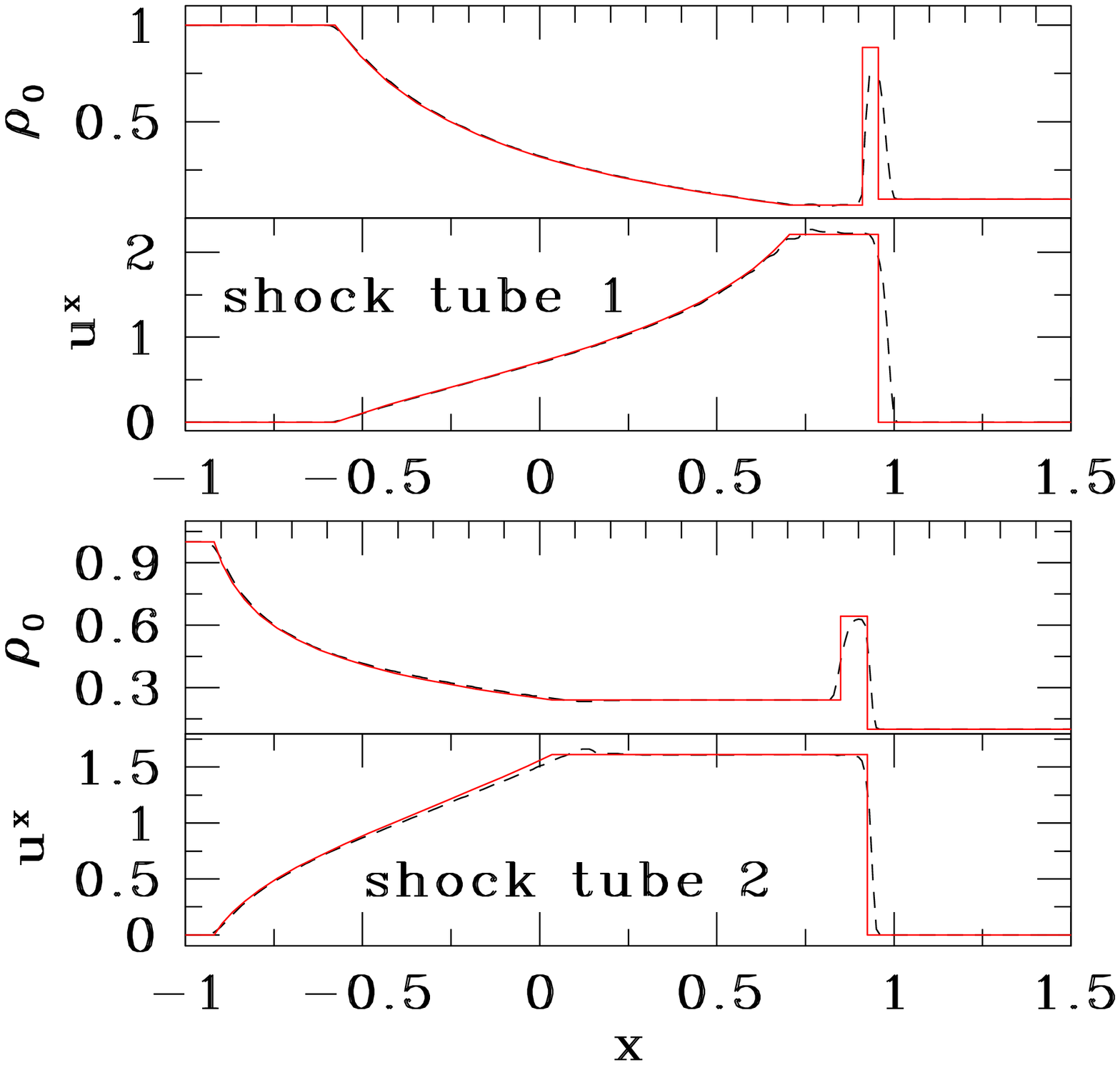}
\caption{Same as Fig.~\ref{fig:fast_slow_shock} but for the 1D shock tube 1 
and shock tube 2 tests.}
\label{fig:shock_tubes}
\end{figure}

We integrate the MHD equations from $t=0$ to $t=t_{\rm final}$, where
$t_{\rm final}$ is specified in
Table~\ref{tab:1Dtests} for each case. The gas satisfies a $\Gamma$-law EOS
with $\Gamma=4/3$, and is evolved on a uniform resolution grid with
$\Delta x = 0.01$.  We are able to integrate all the cases using the PPM reconstruction
scheme and the HLL approximate Riemann solver with a timestep $\Delta t=0.5 \Delta x$. 
We use ``copy'' boundary conditions (i.e.\ hydrodynamic variables are copied 
and the vector potential is linearly extrapolated to the boundary points) in 
all cases. The first 6 tests in Table~\ref{tab:1Dtests} start with discontinuous
initial data at $x=0$, with homogeneous profiles on either side. 
Figures~\ref{fig:fast_slow_shock}--\ref{fig:shock_tubes} show the profiles 
of $\rho_0$ and $u^x$ at time $t=t_{\rm final}$. Notice that the numerical 
results agree very well with the analytic solution in all cases. The overall performance 
of the new MHD scheme in these tests is about as good as our old 
constrained-transport scheme presented in~\cite{dlss05}.

\begin{figure}
\includegraphics[width=9cm]{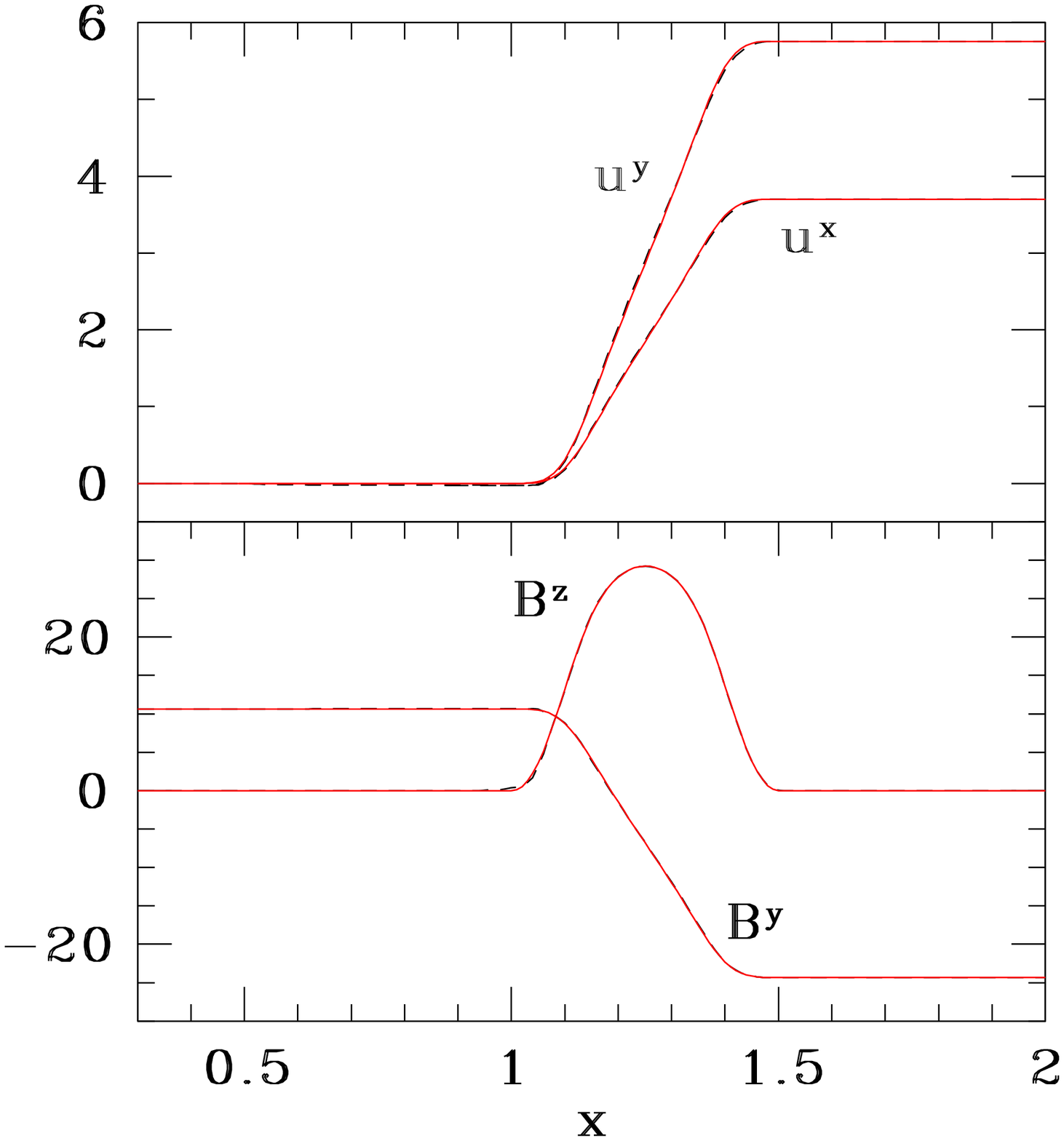}
\caption{1D nonlinear Alfv\'en wave test: MHD variable profiles. Test
  results with resolution $\Delta x = 0.01$ (dashed,
  black lines) are compared to the exact solution (solid, red lines) 
at time $t=t_{\rm final}=2.0$. Our computational domain is $x\in(-2,2)$.}
\label{fig:alfven}
\end{figure}

\begin{figure}
\includegraphics[width=8cm]{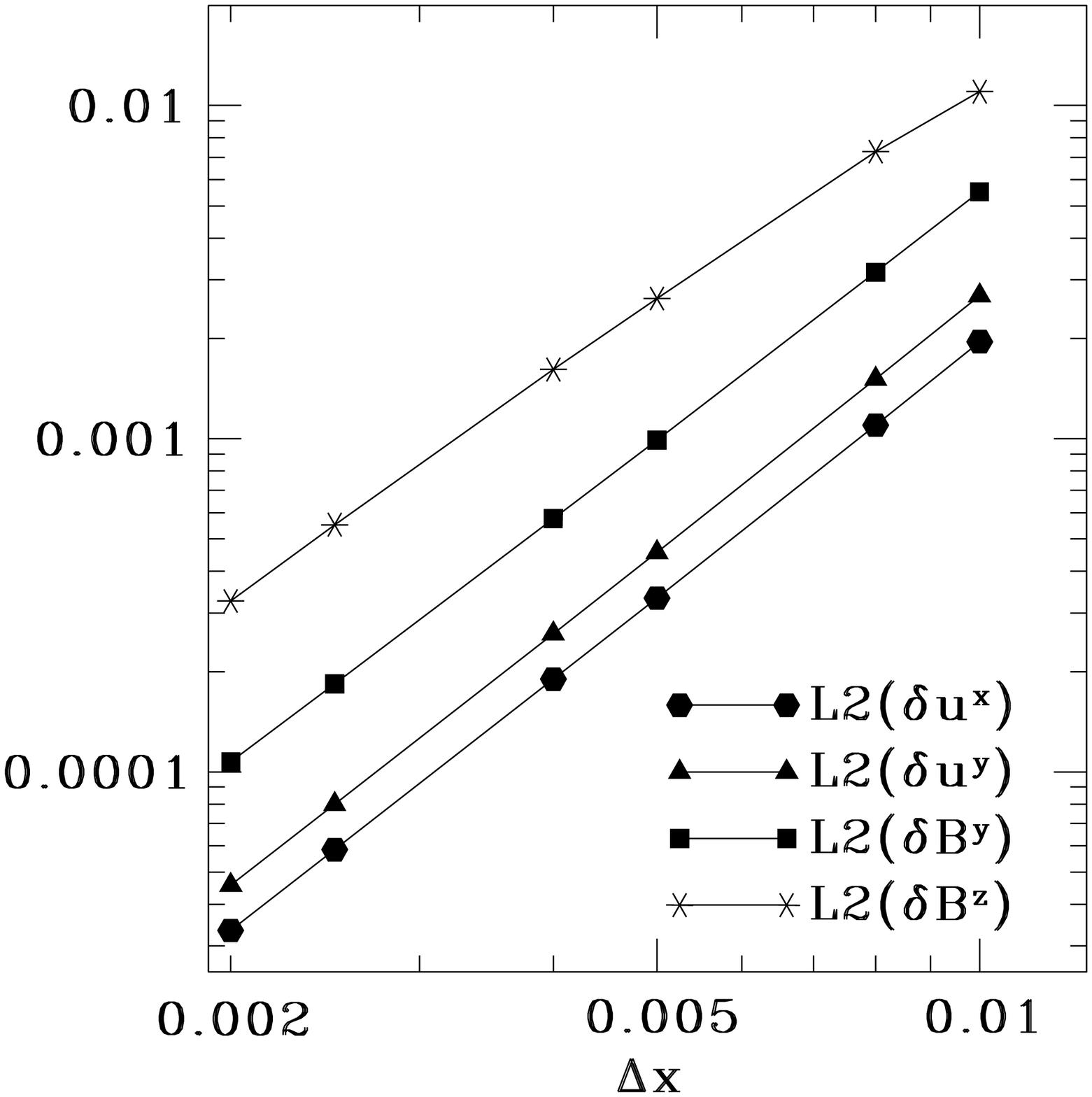}
\caption{1D nonlinear Alfv\'en wave test: L2 norms of the errors in
  $u^x$, $u^y$, $B^y$ and $B^z$ at $t=t_{\rm final}=2.0$. This log-log 
plot demonstrates that L2 norms of the errors are proportional to 
$\Delta x^2$, and are thus second-order convergent.}
\label{fig:alfven_l2}
\end{figure}

In the nonlinear Alfv\'en wave test, unlike the first 6 tests, the two states 
listed in Table~\ref{tab:1Dtests} are joined by a continuous function. 
We use the same initial data described in Appendix~B of~\cite{dlss05}. 
Figure~\ref{fig:alfven} demonstrates very good agreement between numerical 
results and the analytic solution for velocity and magnetic field profiles 
at time $t=t_{\rm final}=2$.
Figure~\ref{fig:alfven_l2} shows the L2 norm of the error in $u^x$, 
$u^y$, $B^y$ and $B^z$, varying only numerical resolution. The L2 norm of a grid function 
$\delta g \equiv g-g^{\rm exact}$ is computed by summing over every 
grid point $i$:
\beq
  L2(\delta g) = \sqrt{\sum_{i=1}^{N} [\delta g(x_i)]^2 \Delta x} \ ,
\eeq
where $N\propto 1/\Delta x$ is the number of grid points. We find that 
the errors converge to zero at second order in $\Delta x$, as expected. 

\subsubsection{Two-dimensional tests} 
\label{sec:2d-min-tests}

We perform the two-dimensional cylindrical blast explosion test 
and rotating disk test described in~\cite{bal_sp99,zbl03}. In both 
tests, all variables are functions of $x$ and $y$ only,  
the initial magnetic field is uniform and oriented along the 
$x$-direction, and the initial velocity does not have the $z$-component. 
Such a uniform magnetic field can be derived from the 
vector potential 
\beq
  A_x=A_y = 0 \ \ , \ \ A_z = y B^x \ .
\eeq

It can be shown from the MHD evolution equations that $A_x=A_y=B^z=v^z=0$ 
remains true for all time $t$.  It can also been shown from the 
finite-volume equations that our MHD evolution scheme preserves 
this property. Our numerical simulations also confirm that 
$A_x=A_y=B^z=v^z=0$ is satisfied to roundoff error at all times. 
It follows from $A_x=A_y=B^z=0$ and $\ve{B}=\ve{\nabla}\times \ve{A}$
that $B^i \partial_i A_z=0$. Hence contours of constant $A_z$ coincide
with the magnetic field lines. 
The evolution of magnetic field thus reduces to the evolution 
of $A_z$, which can be shown to satisfy the simple advection equation: 
\beq
\partial_t A_z + v^i \partial_i A_z = 0 \ .
\label{eq:Azevol}
\eeq
It follows from Eq.~(\ref{eq:Azevol}) 
that the constant $A_z$ contours are comoving with the fluid. 
We note that we do not evolve Eq.~(\ref{eq:Azevol}) 
directly. Instead, we evolve $A_z$ using the HRSC 
scheme described in Sec.~\ref{sec:amr-ct}. 
Small numerical resistivity inherent in our HRSC scheme 
could cause small violations of Eq.~(\ref{eq:Azevol}), 
especially in regions of strong shocks. However, 
the relation $B^i \partial_i A_z=0$ is satisfied to truncation 
error at all times. %It is not satisfied to 
%roundoff error because in our scheme the staggerings of $B^i$ 
%and $A_z$ are not designed to preserve the equality.

{ \it Cylindrical blast explosion}

\begin{figure*}
\includegraphics[width=5cm]{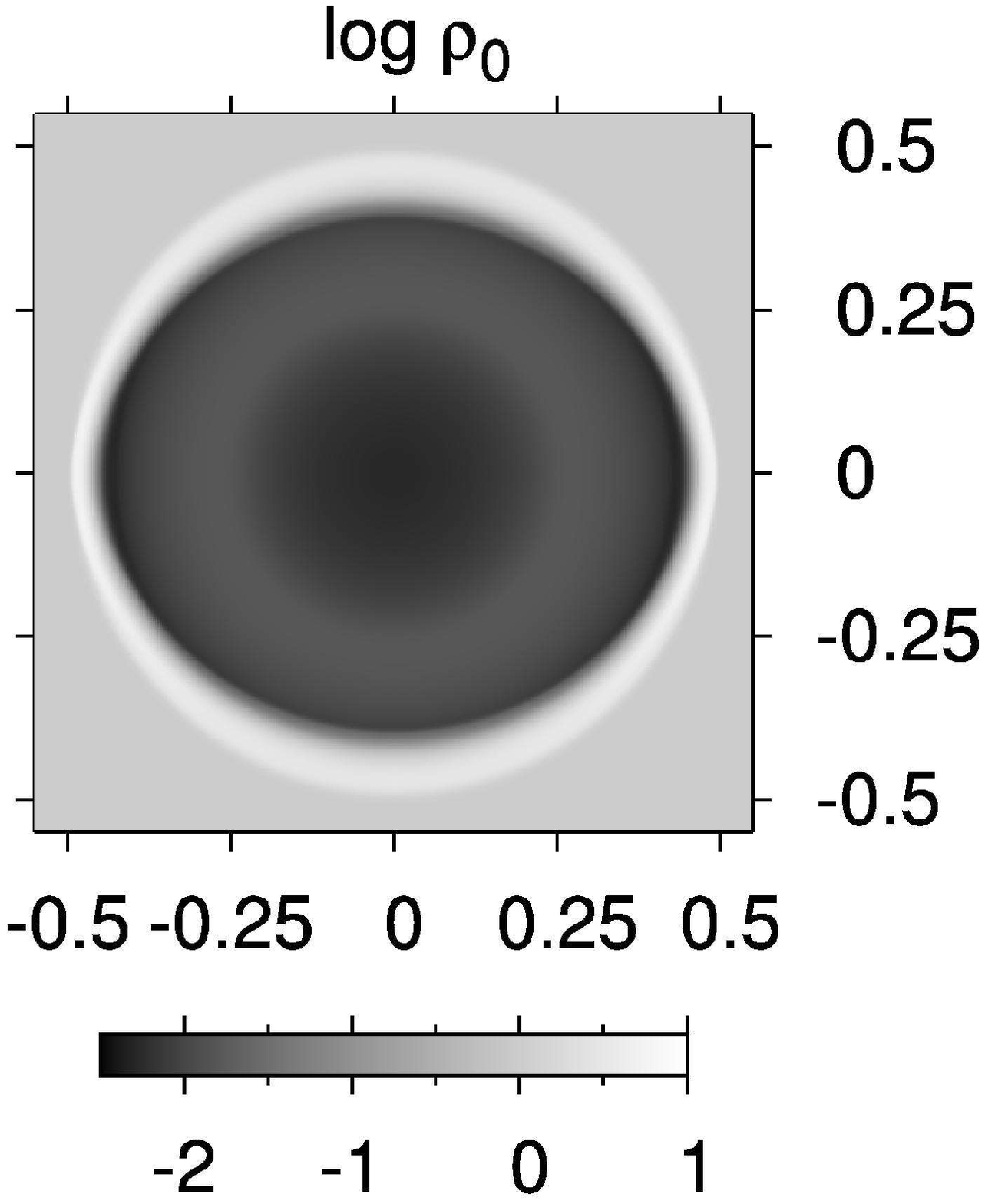}\hspace*{2.2cm}
\includegraphics[width=5cm]{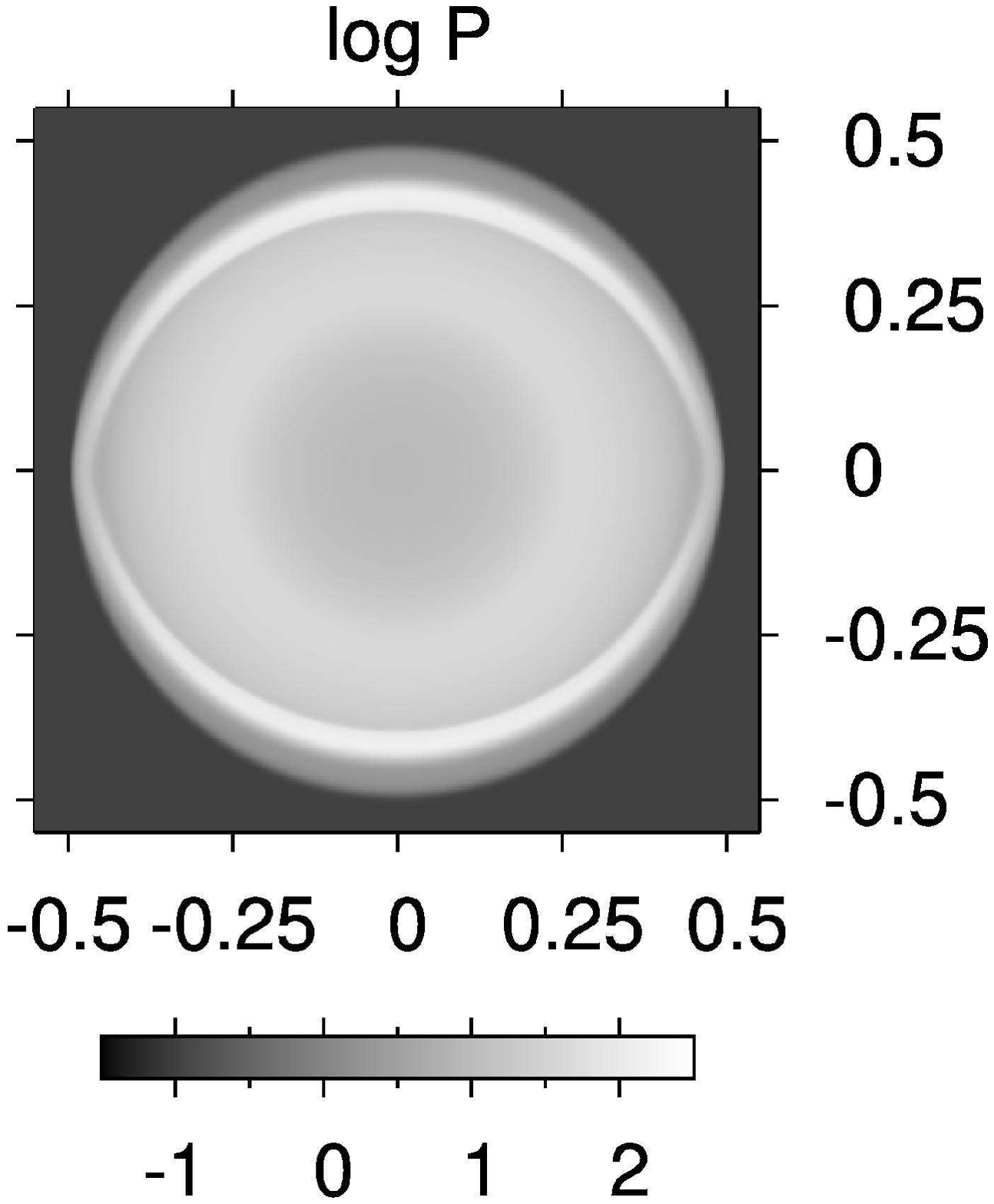}
\vskip 0.5cm
\includegraphics[width=5cm]{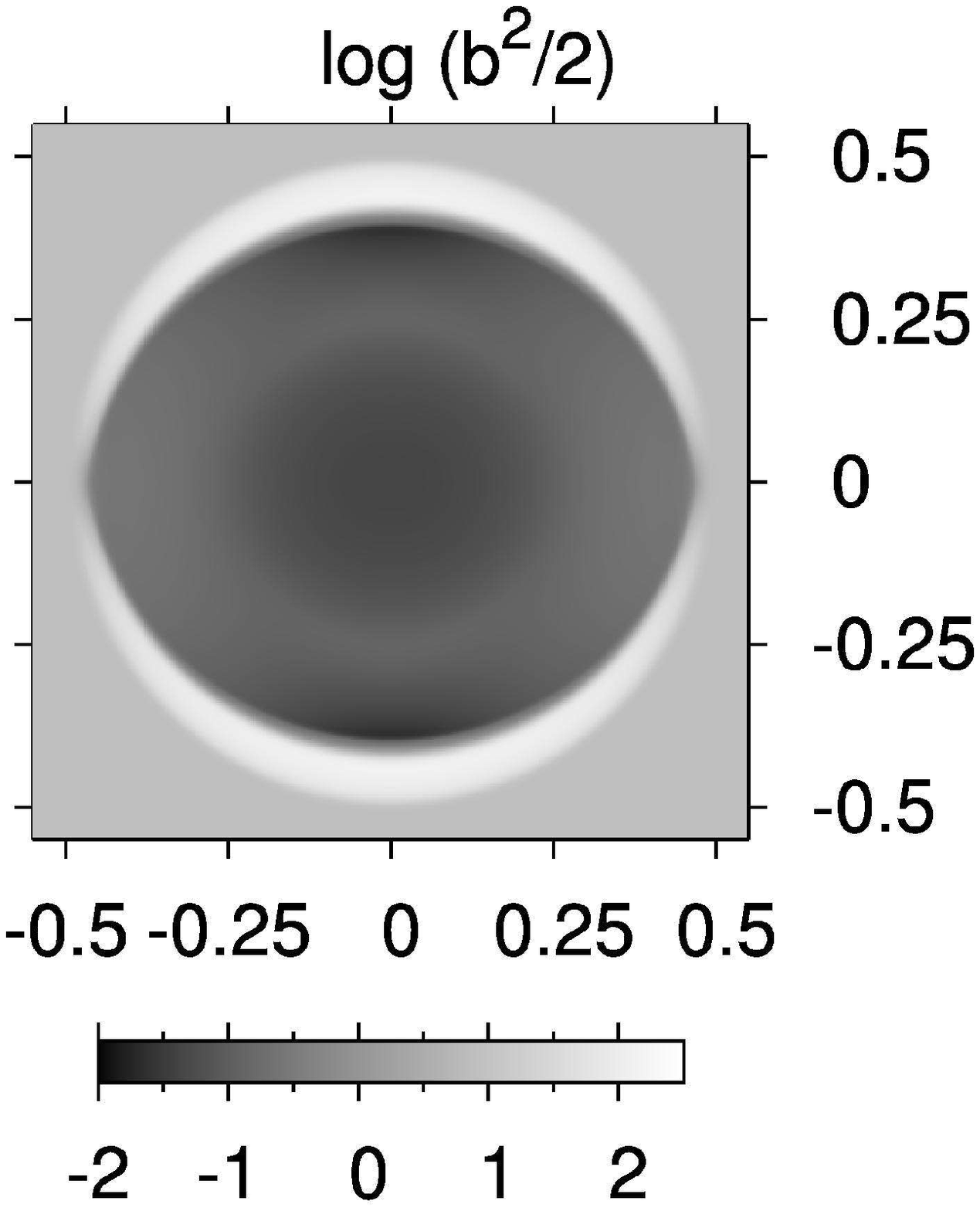}\hspace*{2.2cm}
\includegraphics[width=6cm]{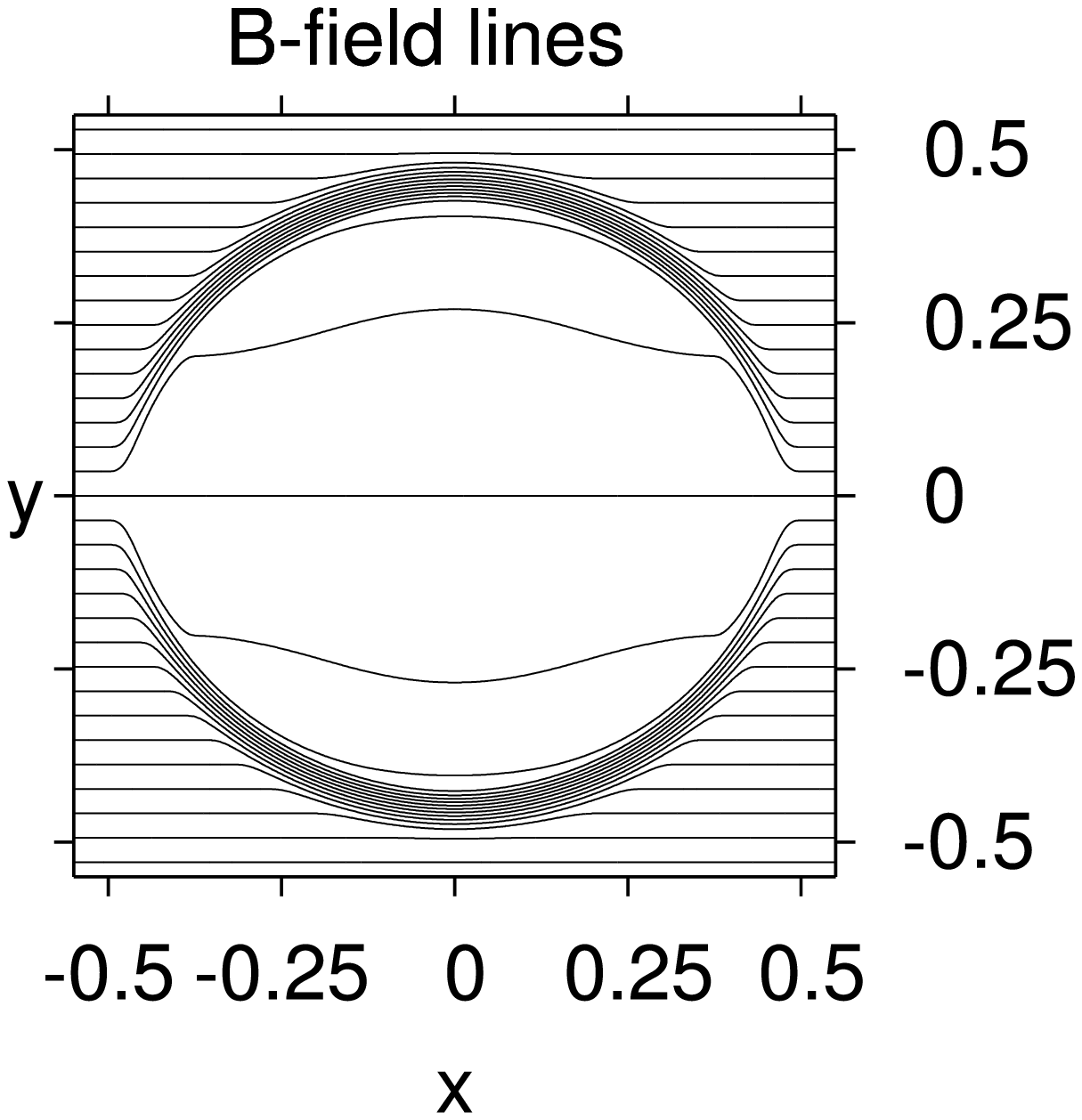}
\caption{Cylindrical blast explosion: 2D MHD variable profiles.
  Density $\rho_0$, gas pressure $P$, magnetic pressure $b^2/2$, and
  magnetic field lines are plotted at $t=0.4$.  The simulation is
  performed with uniform resolution $\Delta x = \Delta y \equiv \Delta = 0.002$. 
Magnetic field lines coincide with contours of $A_z$, and are thus
plotted according to $A_z =0.5i - 8$, with $i=1,2,\dots,31$.}
\label{fig:blast}
\end{figure*}

\begin{figure*}
\includegraphics[width=14cm]{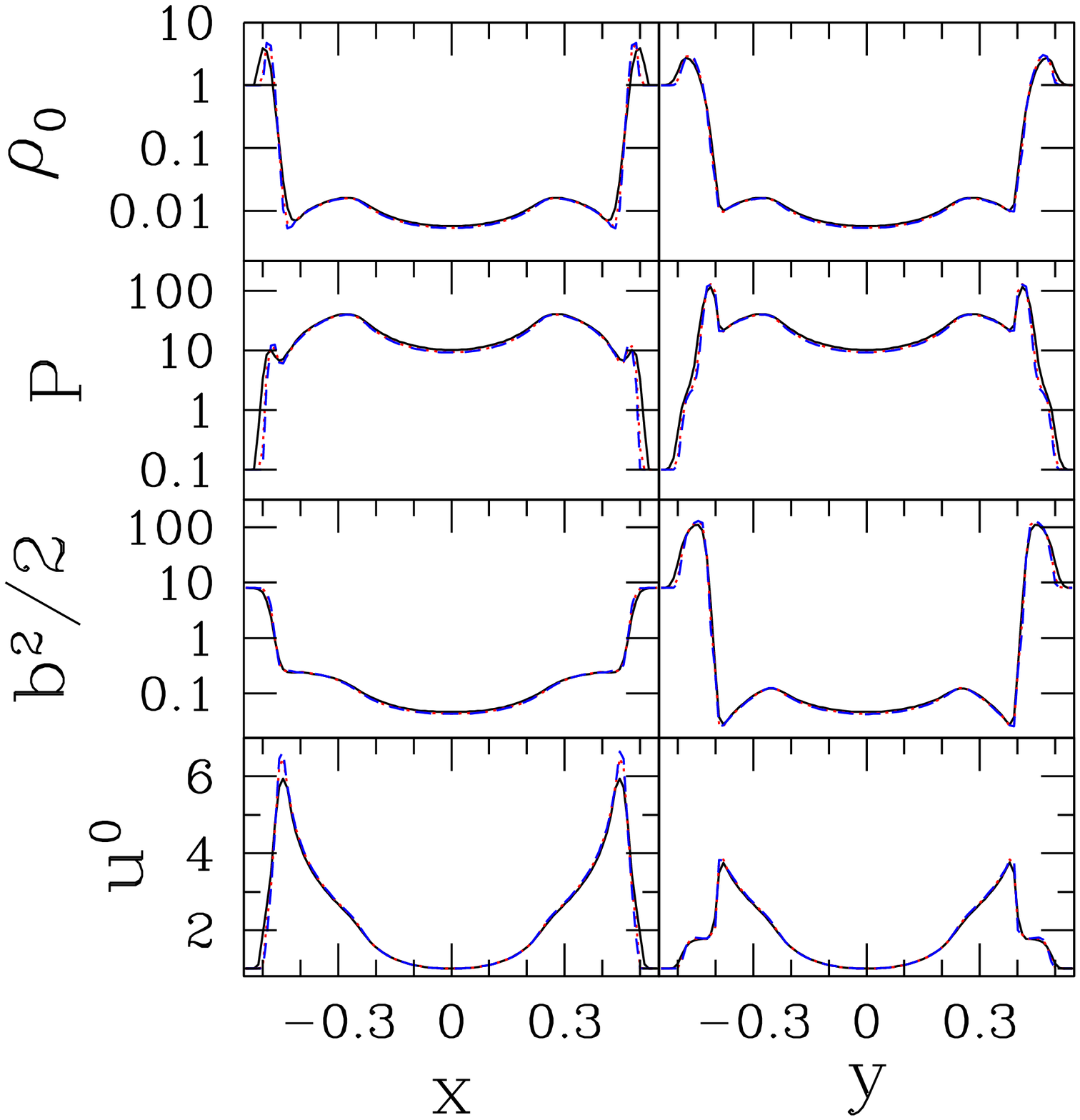}
\caption{Cylindrical blast explosion test: 1D MHD variable profiles
  at different resolutions.  Density $\rho_0$, pressure $P$, magnetic
  pressure $b^2/2$,
  and Lorentz factor $u^0$ profiles along the $x$-axis (left) and
  $y$-axis (right) at $t=0.4$ are plotted at resolutions $\Delta=$0.004 (black
  solid line), 0.0025 (red dotted line) and 0.002 (blue dashed line).}
\label{fig:blast_xy}
\end{figure*}

In this test, the fluid is initially at rest with uniform 
density $\rho_0=1$ throughout the computational domain 
$x\in (-0.55,0.55)$, $y\in (-0.55,0.55)$. Inside a cylinder 
of radius 0.08 is a uniform high pressure $P=10^4$ 
surrounded by an ambient fluid of much lower pressure 
$P=0.1$. The initial magnetic field is $B^x/\sqrt{4\pi}=4.0$, 
and $B^y=B^z=0$ everywhere. The fluid satisfies a $\Gamma$-law 
EOS with $\Gamma=4/3$. We perform simulations with uniform 
resolutions $\Delta x = \Delta y \equiv \Delta=$ 0.004, 0.0025 and 0.002,
applying ``copy'' boundary conditions at the computational domain boundaries. 
We find that evolutions with HLL flux, coupled with either the MC or
PPM reconstruction schemes, result in a code crash due to the strong
initial pressure jump.  
We are able to evolve the system stably by using the minmod 
reconstruction scheme coupled with the LLF flux. A similar finding 
is reported in~\cite{sk05}. 

Figure~\ref{fig:blast} shows the 
two dimensional profile of density $\rho_0$, gas pressure $P$, magnetic pressure 
$b^2/2$ and magnetic field lines at time $t=0.4$, where the blast wave 
has nearly reached the boundary of the computational domain. Figure~\ref{fig:blast_xy} 
shows the one-dimensional profiles along the $x$ and $y$ axis for the three 
resolutions. The profiles are qualitatively similar to the those reported in~\cite{k99,zbl03,sk05}. 
The initial high pressure in the central region causes a strong explosion. 
The explosion is asymmetric in the $x-$ and $y-$directions due to the presence of 
magnetic fields. The explosion is unimpeded in the $x$-direction, so the Lorentz 
factor $u^0$ of the fluid is larger along the $x$-axis than along the $y$-axis, 
as demonstrated in Fig.~\ref{fig:blast_xy}. The magnetic field lines are squeezed 
in the $y$ direction, sapping the magnetic field energy in the central
region, and driving an intense magnetic field in two thin oblate layers surrounding the central 
region (see Fig.~\ref{fig:blast}). By $t=0.4$, the central density and
magnetic pressure have decreased by two orders of magnitude, while the
central gas pressure has dropped by three orders of magnitude.

By comparing the numerical data from the three resolution runs in the entire 
computational domain at $t=0.4$, we see signs of convergence. 
However, the convergence rate is less than first order. This is likely due to the fact 
that the initial strong pressure discontinuity requires resolutions higher than those 
used in our simulations to exhibit the proper convergence, as pointed out 
in~\cite{sk05}. However, we find that in the central $|x|<0.3$,
$|y|<0.3$ region, $\rho_0$, $P$ and $b^2$ converge to second order, while
$u^0$ converges to first order. 

\begin{figure*}
\includegraphics[width=5cm]{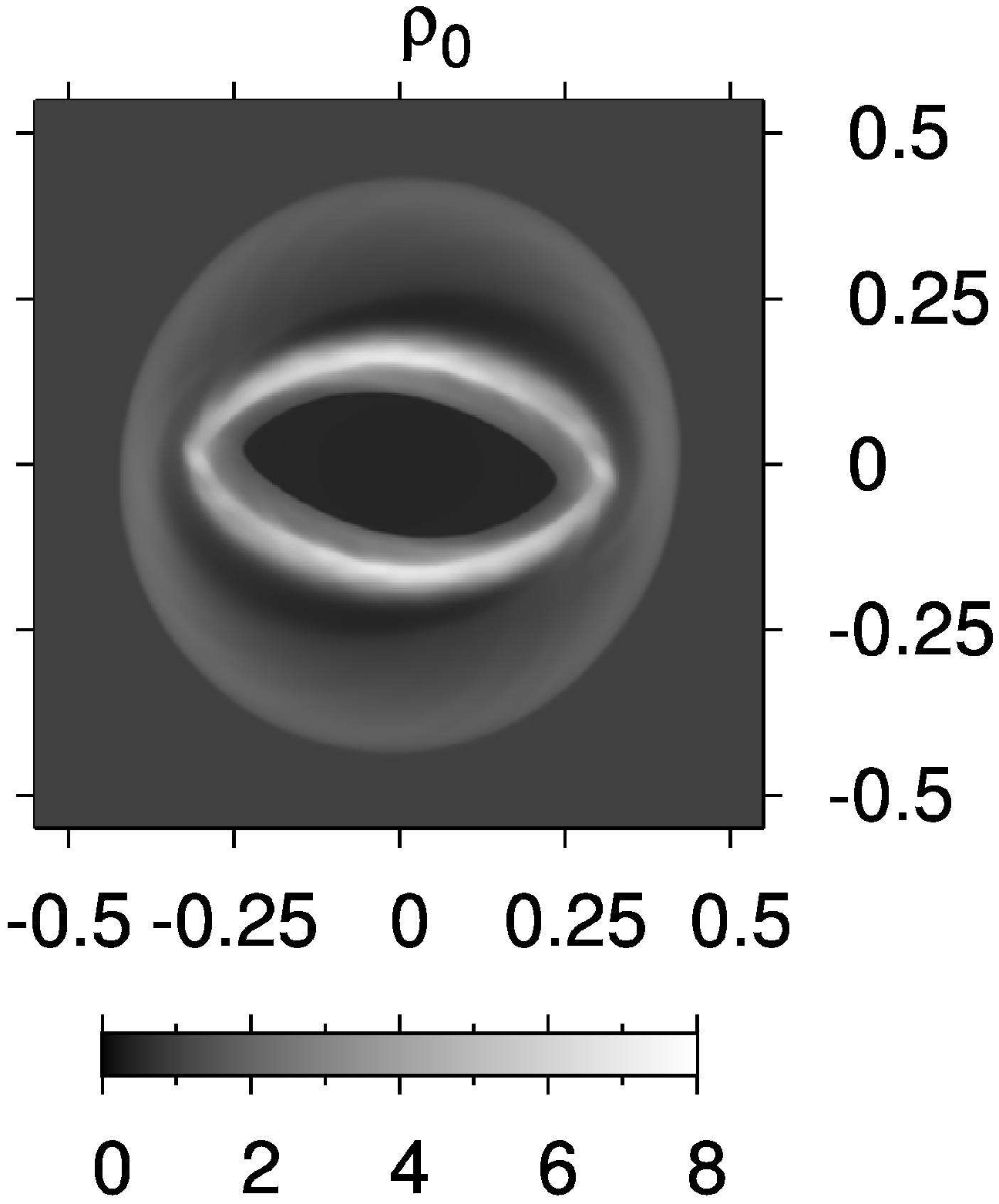}\hspace*{2.2cm}
\includegraphics[width=5cm]{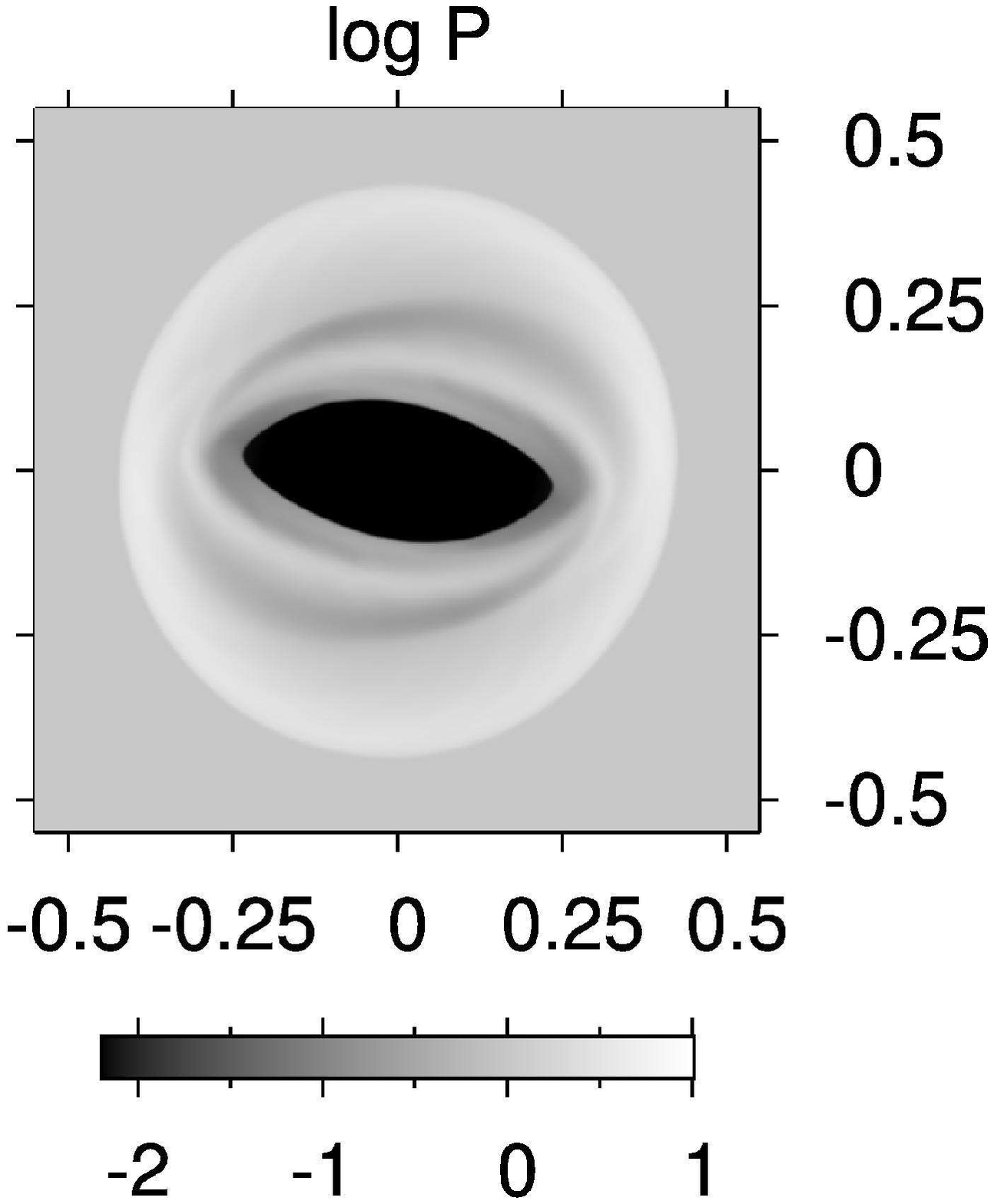}
\vskip 0.5cm
\includegraphics[width=5cm]{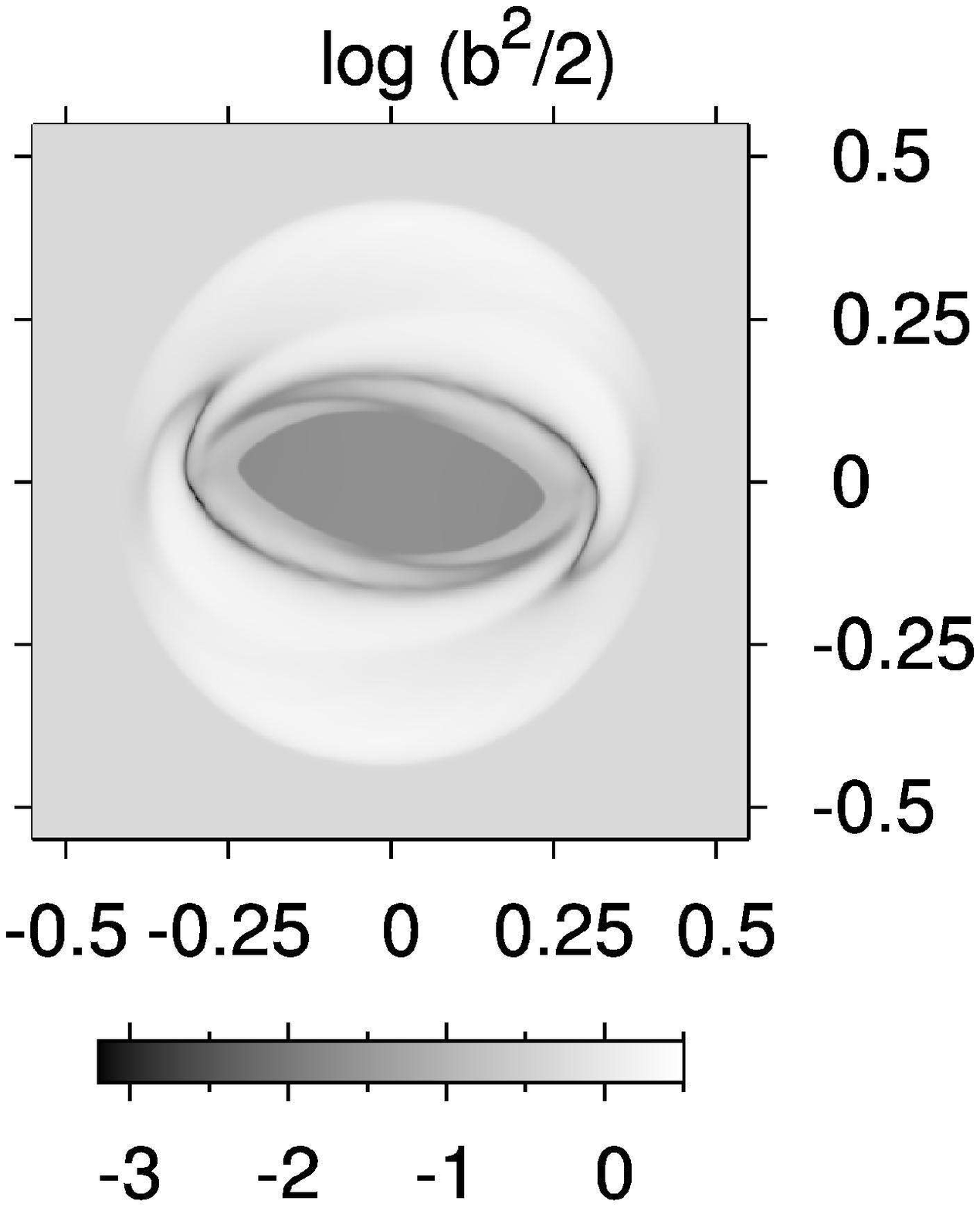}\hspace*{2.2cm}
\includegraphics[width=6cm]{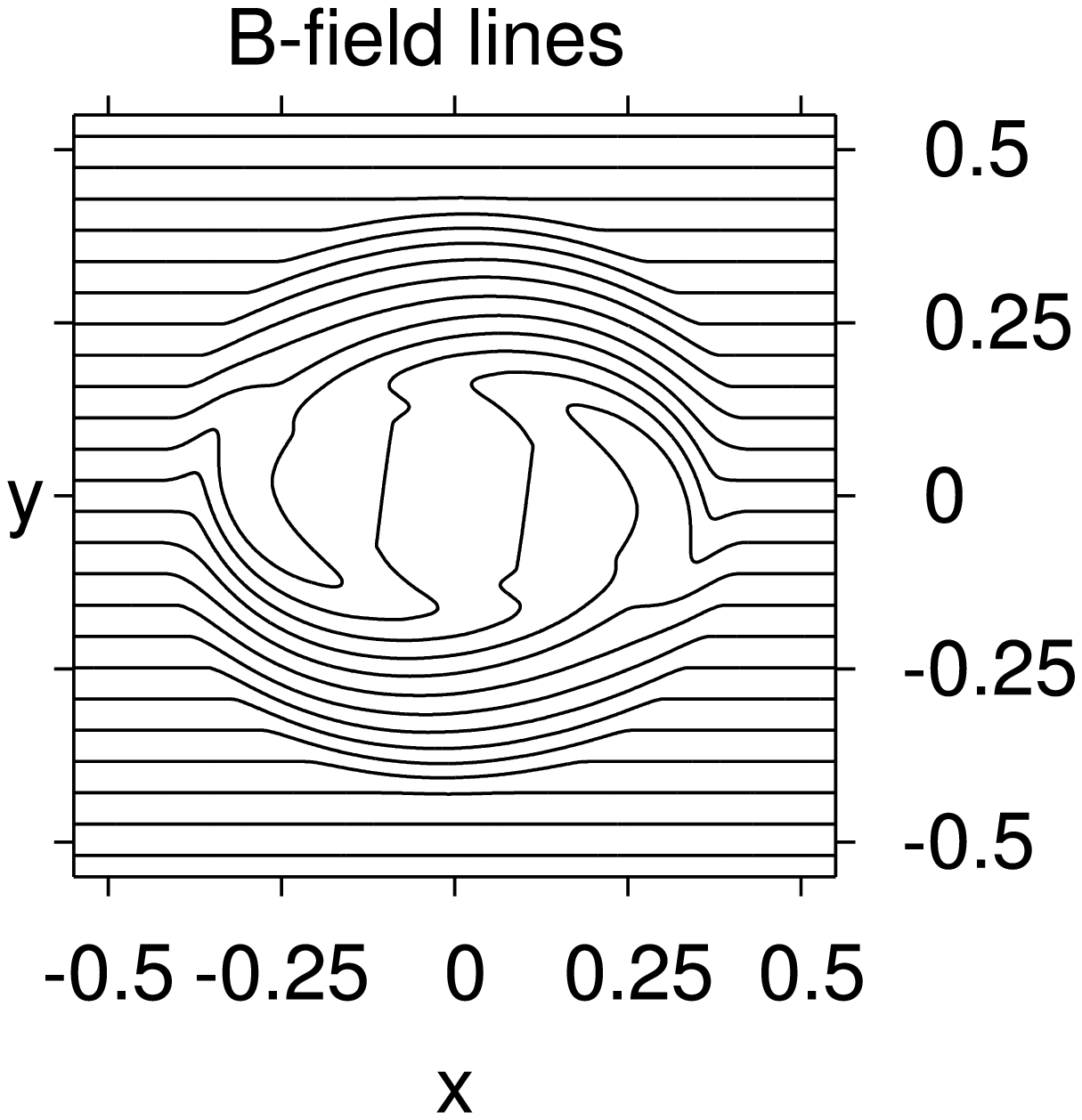}
\caption{Cylindrical rotating disk (rotor) test: 2D MHD variable profiles.
  Density $\rho_0$, gas pressure $P$, magnetic pressure $b^2/2$, and
  magnetic field lines are plotted on the $xy$-plane at $t=0.4$. The
  simulation is performed with a uniform resolution $\Delta x = \Delta y = \Delta =
  0.002$. Magnetic field lines coincide with contours of $A_z$, and
  are thus plotted according to $A_z =0.16 i - 2$, with
  $i=1,2,\dots,24$.}
\label{fig:rotor}
\end{figure*}

\begin{figure*}
\includegraphics[width=14cm]{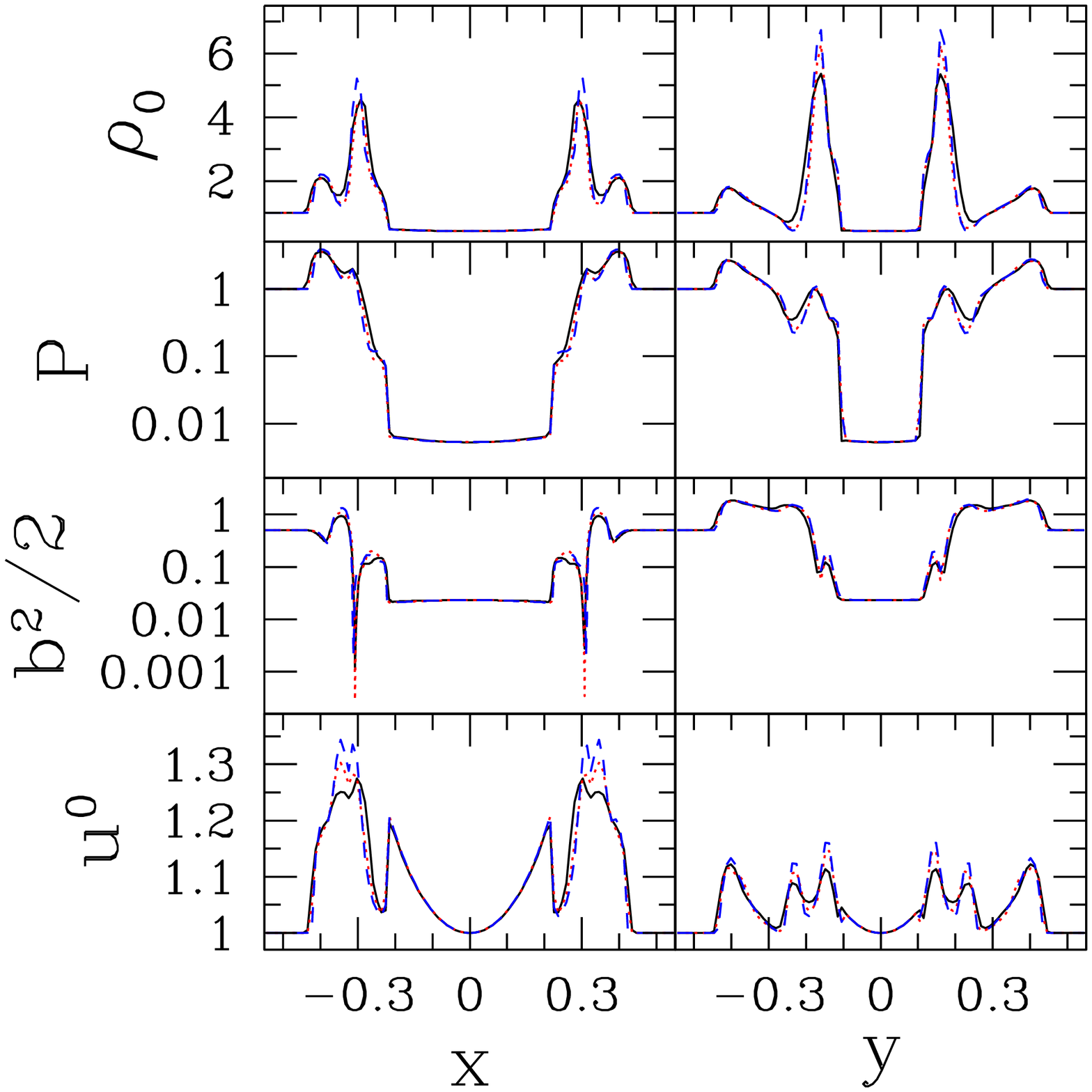}
\caption{Cylindrical rotating disk (rotor) test: 1D MHD variable profiles at
  different resolutions. Density $\rho_0$, pressure $P$, magnetic pressure $b^2/2$, and
  Lorentz factor $u^0$ along the $x$-axis (left) and $y$-axis (right)
  at $t=0.4$ are plotted, at resolutions $\Delta=$0.004 (black
  solid line), 0.0025 (red dotted line) and 0.002 (blue dashed line).}
\label{fig:rotor_xy}
\end{figure*}

{\it Cylindrical rotating disk (rotor)}

The initial configuration of this rotor test consists of a uniform high density 
($\rho_0=10$) central region of cylindrical 
radius 0.1 uniformly rotating with an angular velocity $\omega = 9.95$. The disk 
is surrounded by an ambient gas of density $\rho_0=1$. The gas pressure $P=1$ is 
constant everywhere. The initial magnetic field is uniform and is set to 
$B^x/\sqrt{4\pi}=1$ and $B^y=B^z=0$. The gas satisfies a $\Gamma=5/3$ EOS. 
We evolve the system using the minmod reconstruction scheme coupled with 
the LLF flux and at resolutions $\Delta x=\Delta y =\Delta =$ 0.004, 0.0025 
and 0.002. ``Copy'' boundary conditions are applied at the outer
boundaries for this test.

Figures~\ref{fig:rotor} and \ref{fig:rotor_xy} show the profiles of 
$\rho_0$, $P$, $b^2/2$, $u^0$ and magnetic field lines at time $t=0.4$. 
These profiles are qualitatively similar to those in~\cite{zbl03,sk05}. 
The rotor causes magnetic winding. At time $t=0.4$, the field lines in 
the central region are rotated by $\sim 90^{\circ}$. The winding 
slows down the rotation of the disk. The maximum Lorentz factor decreases 
from the initial value of 10 to 1.7 at $t=0.4$.  
The density, pressure and magnetic 
field in the central region also decrease substantially. A high-density, 
oblate shell is formed surrounding the central region. 

As in the cylindrical explosion test, we see signs of convergence 
as the resolution is increased. However, the overall convergence 
rate is less than first order due to resolutions too low to adequately
resolve the fine structure of the flow. 
The rotor test is even more severe than the cylindrical explosion test. 
This is because the initial Lorentz factor $u^0$ has a steep slope near 
the edge of the disk. Even with our highest resolution $\Delta = 0.002$, 
the initial $u^0$ decreases from 10 at the edge of the disk to 4.5 at the 
next grid point inside the disk. While the three simulations produce the 
same qualitative result, proper convergence order is not likely to 
be achieved when this initial steep feature of the velocity is poorly resolved. 
However, we do find approximate second-order convergence in $\rho_0$ and $b^2$ in the 
the region along the $x$-axis with $|x|<0.2$ before the density, pressure 
and magnetic field display a sudden jump (see Fig.~\ref{fig:rotor_xy}). On the 
other hand, $u^0$ and $P$ converge faster than first order but less than second order 
in that region. 

There are several conserved global quantities in two-dimensional Minkowski 
spacetime: 
\beqn
 M &=& \int \int \rho_0 u^0 dx dy = \sum_{i,j} \bar{\rho}_{*ij} \Delta x \Delta y \ , \\ 
 E &=& \int \int T^{00} dx dy = M + \sum_{i,j} \bar{\tilde{\tau}}_{ij} \Delta x \Delta y \ , \\ 
 P_k &=& \int \int T^0{}_k dx dy = \sum_{i,j} (\bar{\tilde{S}}_k)_{ij} \Delta x \Delta y \ , \\
 J &=& \int \int (x T^0{}_y - y T^0{}_x) dx dy \ ,
\eeqn
where the sum is over all the grid points and the volume average is equivalent to the 
surface average over a grid cell in the $x$-$y$ plane in two dimensions. Since there 
is no source term in Minkowski spacetime [i.e.\ $\ve{S}=0$ in Eq.~(\ref{eq:coneq})], 
our finite-volume scheme should conserve $M$, $E$, and $P_k$ to roundoff error, 
provided that no material flows through the boundary of the computation 
domain (i.e.\ $\ve{F}=0$ at the outer boundary).  This condition is
satisfied in our rotor test, since the ambient medium is static and
the torsional Alfv\'en wave generated by the rotor and the expansion
of the high density gas have not reached the boundary at the end of
our simulations at $t=0.4$. Our numerical data confirm that $M$, $E$
and $P_k$ are indeed conserved to roundoff error. On the other hand,
the angular momentum will not be conserved to roundoff error since we
use Cartesian coordinates to evolve the system. Strict numerical
conservation of angular momentum can be achieved if cylindrical
coordinates are adopted (however, $P_x$ and $P_y$ will not be strictly
conserved in cylindrical coordinates). We find that for the rotor test
at $t=0.4$, $J$ is changed by 1.7\% from its initial value when
evolved with resolution $\Delta=0.004$, 1.2\% with $\Delta=0.0025$ and
1.0\% with $\Delta=0.002$. The slow decrease in $J$ violation with
resolution is again related to the insufficient resolution to resolve
the initial steep $u^0$ profile near the edge of the rotor. We find
that the numerically computed {\it initial} $J$ deviates from the analytic
value by 6.8\%, 2.7\% and 1.8\% for $\Delta=$0.004, 0.0025, and 0.002,
respectively. This indicates that the thin layer near the edge of the
rotor with high initial $u^0$ has a non-negligible contribution to
$J$. Angular momentum conservation can be improved substantially if
the thin layer is well-resolved. 

It follows from the induction equation $\partial_t \ve{B}+\ve{\nabla}\times \ve{E}=0$ that 
the global quantities
\beq
  Q^k = \int \int B^k dx dy 
\eeq
are conserved as long as $\ve{E}$ vanishes at the boundary. Since we do not evolve 
the volume-averaged $B^i$, but instead Eqs.~(\ref{eq:BxtfromA})--(\ref{eq:Azdot}), 
our scheme does not conserve $Q^k$ to roundoff error. 
Instead, the quantities
\beq
  \begin{array}{l}
  Q^x_* = \sum_{ij} \langle B^x \rangle_{i+\half,j} \Delta x \Delta y  \\ 
  \\
  Q^y_* = \sum_{ij} \langle B^y \rangle_{i,j+\half} \Delta x \Delta y 
  \end{array}
\eeq
are strictly conserved in our scheme. Our numerical data confirm this expectation. 
The deviation between $Q^k$ and $Q^k_*$ converges to zero at second order 
with increasing resolution. Unlike the angular 
momentum, the strict conservation of $Q^k_*$ means that $Q^k-Q^k_*$ is 
time independent and therefore will not grow with time during the evolution. 

\subsection{Curved spacetime test: Relativistic Bondi flow} 

Next, we test the ability of our code to accurately evolve the
relativistic MHD equations in a strongly curved spacetime near a
black hole. We perform the magnetized relativistic Bondi accretion test. 
Bondi accretion refers to spherically symmetric, steady-state accretion of 
a unmagnetized, adiabatic gas onto a stationary star.  The gas is assumed to be homogeneous 
and at rest far from the star and flow adiabatically with a $\Gamma$-law EOS. 
Analytic solutions for Bondi accretion onto a Schwarzschild black hole 
are given in~\cite{michel72,shapiro_book_83}. It has been shown that the 
relativistic Bondi solution is unchanged in
the presence of a divergenceless radial magnetic field~\cite{dVh03}. 

This test is a powerful one, since it combines strongly curved 
spacetime and relativistic flows, with an analytic solution against
which we compare our numerical results.  It can also be used to test the ability
of our AMR GRMHD scheme to handle the black hole interior, especially
the coordinate singularity at the center. The use of refinement boxes
is natural, since higher resolution is required in
the vicinity of the black hole, whereas a relatively low resolution is
sufficient to resolve the region far away from the black hole. In
addition to simulations on a fixed background spacetime, we also
evolve the black hole spacetime using the puncture technique. The
spatial metric then evolves from the puncture initial data to the
trumpet solution~\cite{hhpbm07,hhobm08}. 
Although this evolution is a pure gauge effect, the
spatial metric and extrinsic curvature change with time, and the gas
and magnetic field will respond to this change. 

In general, a spherically symmetric spatial metric can be written in the form 
\beq
  {}^{(3)}ds^2 = \Lambda(r,t) dr^2 
+ \lambda(r,t) r^2 (d\theta^2 + \sin^2\theta d\phi^2) \ .
\label{eq:spmetric}
\eeq
It is easy to show that any divergenceless, radial magnetic field is given by
\beq
  B^r(r,t) = \frac{B_0 M^2}{\sqrt{\Lambda(r,t)}\, \lambda(r,t) r^2} \ ,
\label{eq:Br-radial}
\eeq
where $M$ is the mass of the black hole and $B_0$ is a constant characterizing 
the strength of the magnetic field. Cartesian coordinates can be constructed from 
the usual transformation: $x=r\sin\theta \cos \phi$, $y=r\sin\theta \sin\phi$ 
and $z=r\cos\theta$. The Cartesian components of the magnetic field $B^i$ 
is given by 
\beq
  B^i(\ve{x},t) = \frac{B_0 M^2 x^i}{\sqrt{\gamma(\ve{x},t)}\, r^3} \ ,
\label{eq:div_radial_B}
\eeq
where the determinant $\gamma$ of the spatial metric $\gamma_{ij}$ in
Cartesian coordinates is given by $\gamma(\ve{x},t)
=\Lambda(r,t)\lambda^2(r,t)$. It is easy to show that this magnetic
field can be derived from the vector potential 
\beq
  A_x = -\frac{B_0 M^2 y}{r(r+z)} \ \ , \ \ A_y = \frac{B_0 M^2 x}{r(r+z)} \ \ , \ \ A_z = 0 \ .
\label{eq:Aimagbondi}
\eeq
We note that Eq.~(\ref{eq:div_radial_B}) is quite general. The radial coordinate 
$r$ can be the Kerr-Schild radius, the shifted Kerr-Schild radius considered below, 
the isotropic radial coordinate in the puncture initial data, or the radial coordinate 
in the trumpet solution of a Schwarzschild black hole. During the puncture evolution 
of a Schwarzschild black hole, the radial coordinate $r=\sqrt{x^2+y^2+z^2}$ 
in the numerical simulation changes 
from the isotropic radial coordinate to the radial coordinate of the trumpet 
solution. Equation~(\ref{eq:div_radial_B}) will remain valid if the 
evolution preserves spherical symmetry.

The parameters of the magnetized Bondi test presented here are the
same as those used by~\cite{hsw84,dVh03,HARM}. The sonic radius exists at Schwarzschild (areal)
radius $r_s = 8M$. The density is normalized so that the mass accretion rate 
is $\dot{M}=1$, and the equation of state is $\Gamma = 4/3$. The initial data 
for the hydrodynamic variables are given by the analytic solution, and the 
magnetic vector potential is given by Eq.~(\ref{eq:Aimagbondi}). We parametrize 
the strength of the magnetic field by the ratio $b^2/\rho_0$ at the 
event horizon. The relationship between $B_0$ and \bsrho
can be computed analytically and is given by 
\beq
  B_0 = \frac{2.2688}{M} \sqrt{ \left( \frac{b^2}{\rho_0}\right)_{\rm horizon}} 
\label{eq:B0}
\eeq
for the hydrodynamic setup chosen here. We note that even though Eq.~(\ref{eq:B0}) 
is computed in Kerr-Schild radial coordinates, it applies to any other 
radial coordinate because $B_0$ is gauge-invariant. To see this, we compute 
$b^2$ at spatial infinity, where the gas is static. Using Eqs.~(\ref{eq:spmetric}), 
(\ref{eq:Br-radial}) and (\ref{eq:bmu}) we obtain 
\[
  b^2(r) = \gamma_{rr} \frac{(B^r)^2}{4\pi}= \frac{B_0^2 M^4}{4\pi r_a^4} 
\mbox{  for $r\rightarrow \infty$,}
\]
where $r_a = \sqrt{\lambda}\, r$ is the areal radius and we have used the 
fact that $\alpha u^0=\sqrt{1+\gamma^{ij}u_i u_j}=1$ for a static
($u_i=0$) gas.  Hence we can write 
\beq
  B_0 = \lim_{r_a\rightarrow \infty} 4\pi \left( \frac{r_a}{M}\right)^4 b^2(r_a) \ ,
\eeq
which is manifestly gauge-invariant. 

In all of our simulations, we use five refinement boxes with half-side 
lengths of $3.125M$, $6.25M$, $12.5M$, $25M$, and $50M$. The outermost,
lowest-resolution box possesses half-side length $100M$. We only
evolve the space above and on the equatorial  
plane $z\geq0$. Equatorial symmetry is applied to hydrodynamic 
variables and $(r+z)A_i$. All variables at the outer boundary are frozen 
to their initial values. Our standard resolution is 
$\Delta x = \Delta y = \Delta z =\Delta = 2.5M$ in the coarsest level. 
The grid spacing $\Delta$ decreases by a factor of two at each successive 
refinement level, so the resolution on the finest level is $\Delta_{\rm min}=M/12.8$. 
For the purposes of testing convergence, we also perform a simulation
in which the resolution is scaled up so that $\Delta_{\rm min}=M/16$. 
To measure errors due to moving refinement boxes in our AMR scheme, we
move refinement box centers according to
\beq
  x_c = x_m \sin \omega t \ \ , \ \ y_c = y_m (1-\cos \omega t) \ ,
\eeq
where we set the parameters $x_m=1.0M$, $y_m=0.6M$ and 
$2\pi/\omega=50M$. Below, we present results for the 
fixed background spacetime simulations and the puncture evolution. 
Without loss of generality, we set $M=1$ in all of our simulations.

\subsubsection{Fixed background spacetime} 

\begin{figure}
\includegraphics[width=9cm]{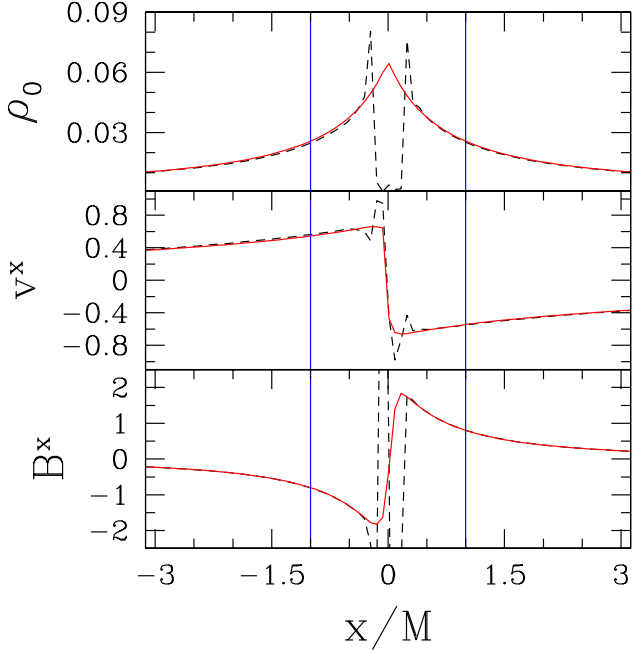}
\caption{Fixed-background, \bsrho=4 magnetized Bondi test: 1D MHD
  variable profiles.  $\rho_0$, $v^x$ and $B^x$ are plotted in the
  equatorial
plane ($z=0$) along the line $y=0.01M$ at $t=101.25M$. Solid 
(red) lines are the analytic solution and dashed (black) lines are
numerical data with $\Delta_{\rm min}=M/12.8$ on the finest
refinement level, using MC reconstruction. The vertical lines denote the 
location of the event horizon $|x|=M$.}
\label{fig:mag_bondi_profiles_mc}
\end{figure}

\begin{figure}
\includegraphics[width=9cm]{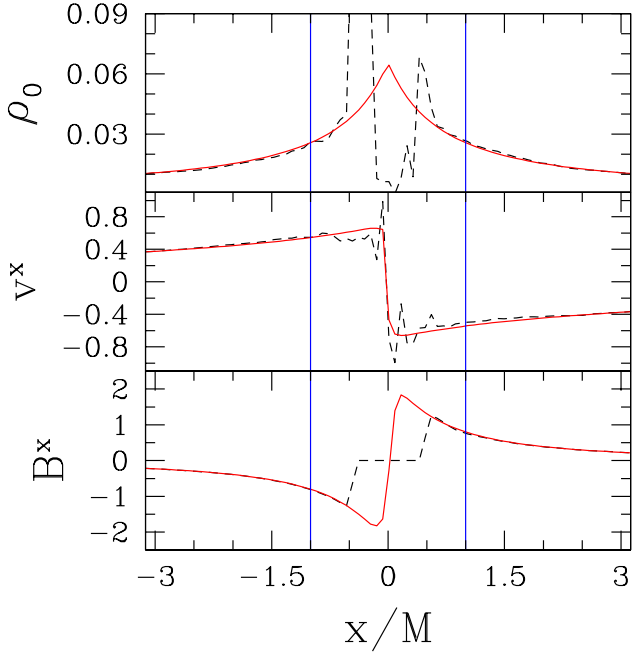}
\caption{Same as Fig.~\ref{fig:mag_bondi_profiles_mc}, but with PPM
  reconstruction for the numerical data, and setting $A_i=0$ for 
  deep inside the BH ($r<0.5M$).}
\label{fig:mag_bondi_profiles_ppm}
\end{figure}

\begin{figure}
\includegraphics[width=9cm]{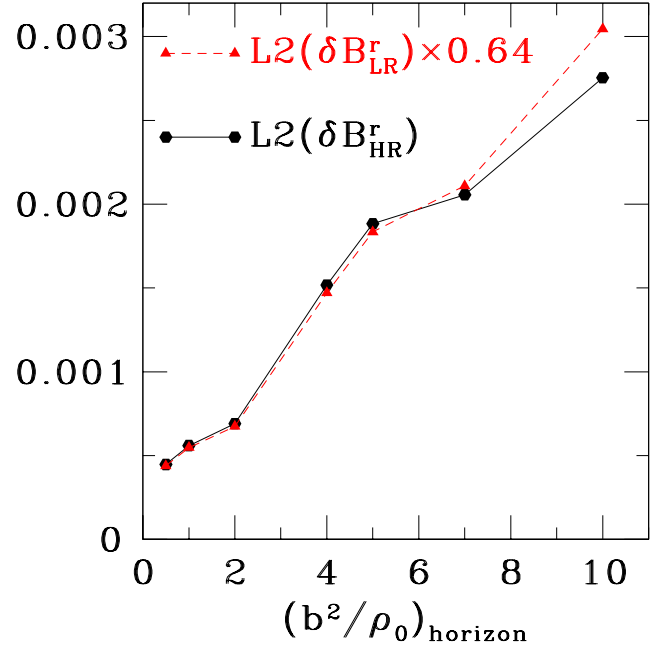}
\caption{Fixed-background magnetized Bondi test: Convergence
  study using MC reconstruction.  L2 norm of $B^r$ as a function of \bsrho is
plotted at $t=101.25M$ for lower ($\Delta_{\rm min}=M/12.8$) 
and higher ($\Delta_{\rm min}=M/16$) resolution runs. The lower resolution result 
is multiplied by the factor 0.64 to demonstrate second-order convergence.}
\label{fig:Br_l2}
\end{figure}

In many relativistic Bondi tests, Kerr-Schild coordinates are used together 
with excision. A different approach is adopted here. We first 
define a shifted Kerr-Schild radius $r=r_{KS}-r_0$, where $r_{KS}$ is the 
Kerr-Schild radius and $r_0$ is a constant 
chosen in the range $0<r_0<2M$. We then construct 
Cartesian coordinates using the standard transformation between 
($x,y,z)$ and ($r,\theta,\phi$). The origin $x=y=z=0$ therefore corresponds 
to a Kerr-Schild radius $r_{KS}=r_0$. The region $r_{KS}<r_0$ is excluded 
in this coordinate system and so is the black hole spacetime singularity. However, 
the origin is a coordinate singularity since the whole surface $r_{KS}=r_0$ 
is mapped to a single point. This coordinate system thus mimics the 
trumpet solution of a Schwarzschild black hole. 
We set $r_0=M$ for all the tests presented in this section. Just as in 
puncture evolutions, we do not use excision but shift the grid 
slightly so that the origin is not on a grid point. 

We are able to evolve the system stably using the MC reconstruction scheme 
coupled with the HLL flux for \bsrho$\lesssim 10$ with our standard resolution. Higher magnetic fields 
may be evolved if the resolution is increased. During the evolution, the 
magnetic field, as well as the density, increases linearly with time 
near the origin, creating jumps in the magnetic field that increase with 
time. This phenomenon causes the evolution near the origin to become more and 
more inaccurate. The inaccurate data spread out slowly from the origin
to the apparent horizon, eventually crossing into the BH exterior. To
overcome this difficulty, we add fourth-order  
Kreiss-Oliger dissipation to the evolved variables inside the horizon 
for radius $r<0.8M$. We also set a density cap $\rho_0<1$ for radius $r<0.5M$. 
This technique stabilizes the evolution near the origin and the system 
quickly settles down to a steady state inside the horizon. 
Figure~\ref{fig:mag_bondi_profiles_mc} shows the profiles of $\rho_0$, 
$v^x$ and $B^x$ in the equatorial plane ($z=0$) along the line $y=0.01$ 
for \bsrho=4 at $t=101.25M$, by which time the center of the refinement boxes has gone 
through slightly more than two rotations. The analytic solution and numerical data are 
plotted together for comparison. Vertical lines denote the location of the horizon 
$|x|=M$. We see that the profiles agree very well 
with the analytic solution outside the horizon. There are strong jumps in the magnitude
and direction of the magnetic field near the coordinate singularity at
the origin. The maximum and minimum values of $B^x$ 
near the origin are 13.4 and -19.5 respectively, far outside the scale shown in the figure. 
However, these jumps are always contained near the coordinate 
singularity. 

The evolution near the coordinate singularity at the origin is less
stable when evolved with PPM reconstruction.  
To remedy this, we set $A_i=0$ for radius $r<0.5M$, well inside the 
horizon, in addition to the technique described
above. Figure~\ref{fig:mag_bondi_profiles_ppm}
shows profiles of MHD variables using PPM. We again see that the profiles agree well 
with the analytic solution outside the horizon, oblivious to the ruggedness of profiles 
in the black hole interior. 

To check for convergence, we perform a number of simulations of varying 
\bsrho with two different resolutions $\Delta_{\rm min}=M/12.8$ and 
$\Delta_{\rm min}=M/16$. We compute the L2 norm of $B^r$ at $t=101.25M$ by 
summing over grid points 
\beq
  L2(B^r) = \frac{\sum (B^r_{\rm numerical}-B^r_{\rm analytic})}{\sum B^r_{\rm analytic,1}} \ ,
\eeq
where $B^r_{\rm analytic,1}$ denote the analytic values of $B^r$ for \bsrho=1. 
We only computed the L2 norm in the innermost refinement level outside the horizon 
with $|x|<3M$, $|y|<3M$, $0\leq z<3M$ and $r>M$. This is the region in the 
black hole exterior where the magnetic field is the strongest. Figure~\ref{fig:Br_l2} 
shows the L2 norm as a function of \bsrho. We see second-order convergence for 
\bsrho$\lesssim 8$. The convergence rate appears to be higher than second-order 
for \bsrho=10, indicating that the data in the lower resolution run
may not be accurate enough to display proper convergence.

\subsubsection{Puncture evolution}

\begin{figure}
\includegraphics[width=9cm]{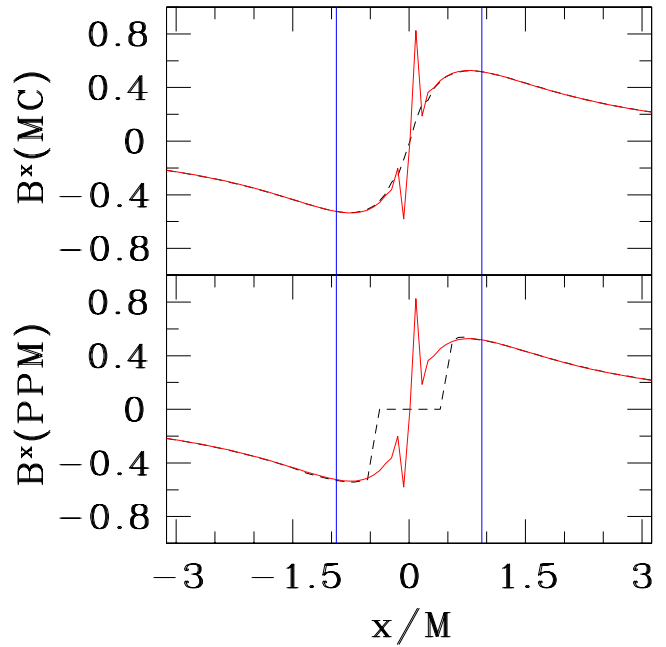}
\caption{Evolved-spacetime, \bsrho=4 magnetized Bondi test: Profile of $B^x$ at 
$t=101.25M$ in the equatorial plane ($z=0$) along the line $y=0.01M$. 
The metric is evolved using the puncture technique. The upper 
graph plots numerical data using MC reconstruction, and the lower graph 
shows the result using PPM reconstruction and setting $A_i=0$ for
$r<0.5M$.
Dashed 
(black) lines are numerical data, and solid (red) lines are results computed by 
Eq.~(\ref{eq:div_radial_B}) with $\sqrt{\gamma}=e^{6\phi}$ taken from 
numerical data. The glitch near $x=0$ results from the loss of accuracy of 
the metric data close to the puncture. The vertical lines denote the location of the black hole 
horizon $|x|=0.94M$.}
\label{fig:mag_bondi_puncture}
\end{figure}

\begin{figure}
\includegraphics[width=9cm]{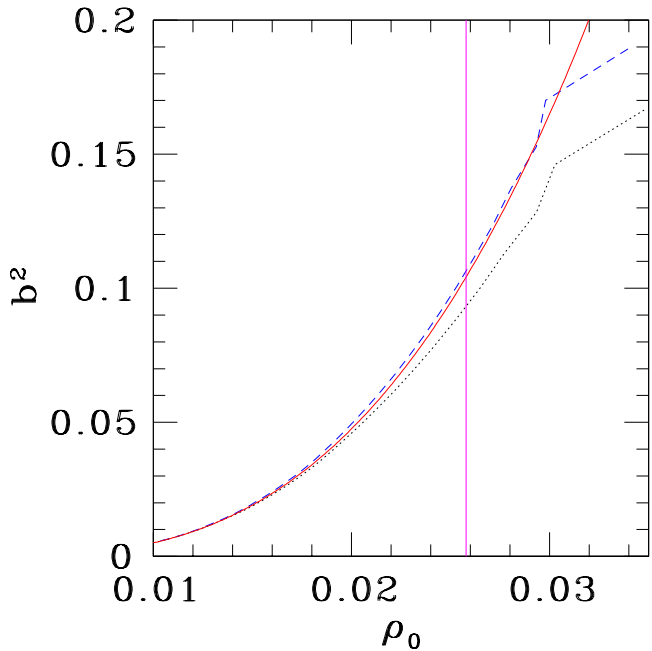}
\caption{Evolved-spacetime, \bsrho=4 magnetized Bondi test: Convergence of $b^2$ as a function of $\rho_0$ at $t=101.25M$
  along the line $x>0.5M$, ($y,z$)=(0,$0.01M$). The background metric is evolved
using the puncture technique. Dotted (black) and Dashed (blue) lines plot 
the numerical data evolved with MC reconstruction using resolutions 
$\Delta_{\rm min}=M/12.8$ and $\Delta_{\rm min}=M/16$, respectively. 
The solid (red) line denotes the analytic profile, and the vertical line
demarcates the horizon boundary, where $\rho_0=(\rho_0)_{\rm
  horizon}=0.02579$. The region with $\rho_0>(\rho_0)_{\rm horizon}$
lies inside the horizon.}
\label{fig:rho_b2_puncture}
\end{figure}

In addition to the evolution with a fixed background metric, we perform several 
magnetized Bondi tests with a {\it time-dependent} background metric. We evolve the 
black hole spacetime using the puncture technique.  In order to 
compare with the analytic solution, we set the matter and EM field source 
terms to zero in the BSSN equations, so that the gas and EM field 
do not affect the spacetime evolution, consistent with assumptions used when deriving  
the analytic solution. However, the gas and EM field will respond to 
the change of the background metric since the MHD and induction equations 
contain metric quantities. The metric evolution is a pure gauge effect: 
the spatial metric and extrinsic curvature evolve from the initial 
maximal, wormhole slicing to the final slicing representing the 
trumpet geometry. 

We evolve the MHD and induction equations using both MC and PPM 
reconstruction, coupled with the HLL flux. A fourth-order Kreiss-Oliger 
dissipation is applied to the MHD evolution variables for $r<0.5M$, which 
is inside the horizon at all times. As before, we set $A_i=0$ for $r<0.5M$ 
in the PPM run to stabilize the evolution near the puncture.
Figure~\ref{fig:mag_bondi_puncture} shows the profile of $B^x$ at $t=101.25M$ 
in the equatorial plane ($z=0$) along the line $y=0.01M$ for \bsrho=4. 
Numerical data are compared to Eq.~(\ref{eq:div_radial_B}) with 
$\sqrt{\gamma}=e^{6\phi}$ taken from the numerical data. We see that 
the data agree well with the analytic result outside the horizon in both runs. 
The glitch near the origin results from the loss of accuracy of $\phi$ 
near the puncture. When compared with Fig.~\ref{fig:mag_bondi_profiles_mc}, 
we see that the $B^x$ profile is smooth in the puncture evolution with MC 
reconstruction. However, we find a similar feature in the $\rho_0$ profile 
as in Fig.~\ref{fig:mag_bondi_profiles_mc}.

Since the analytic solution of the hydrodynamic quantities are given in 
Kerr-Schild coordinates, direct comparison of numerical and analytic 
results is not easy in these simulations. However, since both $b^2$ and 
$\rho_0$ are scalar and 
the system is stationary, the profile of $b^2$ as a function of $\rho_0$ 
is gauge-independent. Figure~\ref{fig:rho_b2_puncture} shows this function 
at $t=101.25M$ in the equatorial plane along the line $y=0.01$ and $x>0.5M$ 
for two resolutions. 
The numerical profile of $\rho_0$ is no longer monotonically increasing 
with decreasing $r$ when the numerical data inside the 
horizon are included, due to inaccuracy near the puncture. We therefore 
remove the data points inside the horizon for $x<x_0$ to prevent multiple 
values of $b^2(\rho_0)$ from appearing in the plot, where $x_0$ is the 
point when $\rho_0$ reaches the maximum. The position of the horizon is 
indicated by the vertical line $\rho_0=(\rho_0)_{\rm horizon}=0.02579$, the 
value of $\rho_0$ at the horizon. The deviation between 
the numerical data and analytic result becomes visible close to the 
horizon in the lower 
resolution run $\Delta_{\rm min}=M/12.8$. Much better agreement is 
achieved in the higher resolution run with $\Delta_{\rm min}=M/16$. 
This is not surprising since $M/12.8$ is a fairly poor resolution for 
puncture simulations. 

\subsection{Curved spacetime test: Collapse of magnetized rotating relativistic star}
\label{sec:bondi} 

This test focuses on magnetized, rotating, relativistic
stellar-collapse simulations. The initial stellar configuration is the
same as Star~D in~\cite{dmsb03} and Star~B in~\cite{dlss05}. The star
satisfies a $\Gamma=2$ polytropic EOS and is uniformly rotating with
$J/M^2=0.34$, where $J$ is the angular momentum. The ADM mass of the
star is $M=1.04M_{\rm TOV}$, where $M_{\rm TOV}$ is the maximum ADM
mass of a non-rotating relativistic star satisfying $\Gamma=2$
EOS. The star is on the unstable branch of the constant $J$ sequence,
and previous numerical simulations have demonstrated that it is dynamically
unstable to gravitational collapse~\cite{dmsb03,dlss05}. 

In all of our simulations, we use seven refinement boxes with half-side 
lengths of $0.9143M$, $1.829M$, $3.657M$, $7.314M$, $14.63M$, $29.26M$
and $58.51M$. The box containing the outer boundary has half-length $117.0M$. The initial 
coordinate radius of the star in the equatorial plane is $3.485M$. Hence 
the stellar interior is initially covered by the three innermost 
refinement boxes. The grid 
spacing is reduced by a factor of two at each successive refinement level. 
We perform three simulations with the resolution in the finest refinement level 
set to $\Delta_{\rm min} = 0.02857M$ (low resolution run), 
$0.02287M$ (medium resolution run), and $0.01829M$ (high resolution run).

We evolve the metric using a fourth-order finite-differencing scheme.
We adopt the puncture gauge conditions with the shift 
parameter $\eta$ set to $0.5/M$. The MHD and induction equations are evolved using 
the PPM reconstruction scheme coupled with the HLL flux. 
Equatorial symmetry is applied to all 
variables. We maintain a low density atmosphere in the computational 
domain with $\rho_{\rm atm}=10^{-10}\rho_{\rm max}(0)$ and 
$P_{\rm atm}=P_{\rm cold}(\rho_{\rm atm})$ as described in 
Sec.~\ref{sec:low-density}.
The Sommerfeld outgoing wave boundary condition is applied 
to the BSSN evolution variables, and outflow boundary conditions are 
applied to the hydrodynamic primitive variables, while the vector potential 
$A_i$ is linearly extrapolated to the boundary. 

\begin{figure}
\includegraphics[width=9cm]{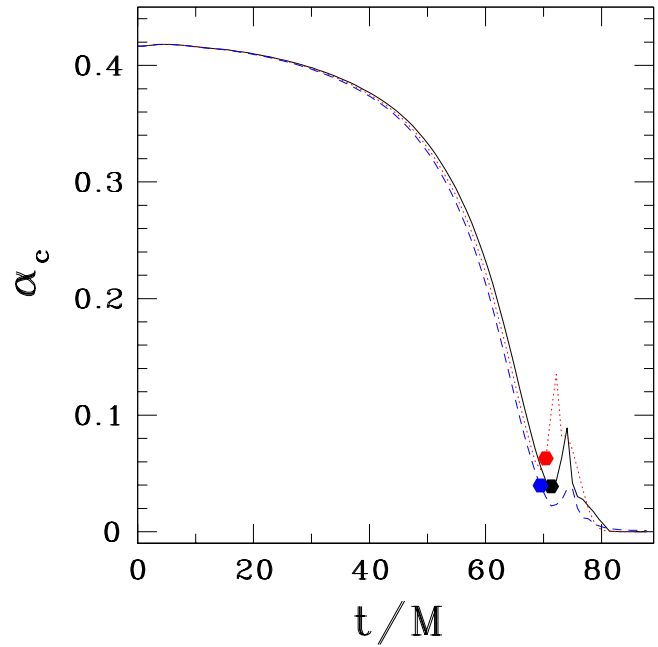}
\caption{Magnetized stellar collapse test: Evolution of the central lapse $\alpha_c$ for the low (black solid line), 
medium (red dotted line), and high (blue dashed line) resolution runs. 
The dot in each case indicates the time at which the apparent horizon appears. 
The increase in $\alpha_c$ soon after the horizon formation is caused by the loss 
of accuracy in metric evolution near the newly formed puncture, which is located 
near the coordinate origin and is deep inside the horizon.}
\label{fig:alphac}
\end{figure}

\begin{figure}
\includegraphics[width=9cm]{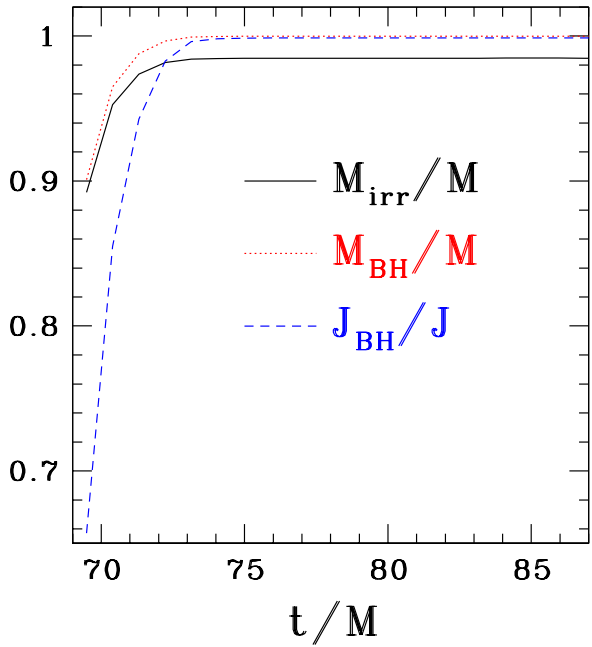}
\caption{Magnetized stellar collapse test: Evolution of the irreducible mass $M_{\rm irr}$, black hole mass $M_{\rm BH}$ and 
angular momentum $J_{\rm BH}$, as normalized by the initial ADM mass $M$ and 
angular momentum $J$. Shown here are data from the high resolution run. 
Results from the low and medium resolution runs are similar.}
\label{fig:horizon}
\end{figure}

\begin{figure}
\includegraphics[width=9cm]{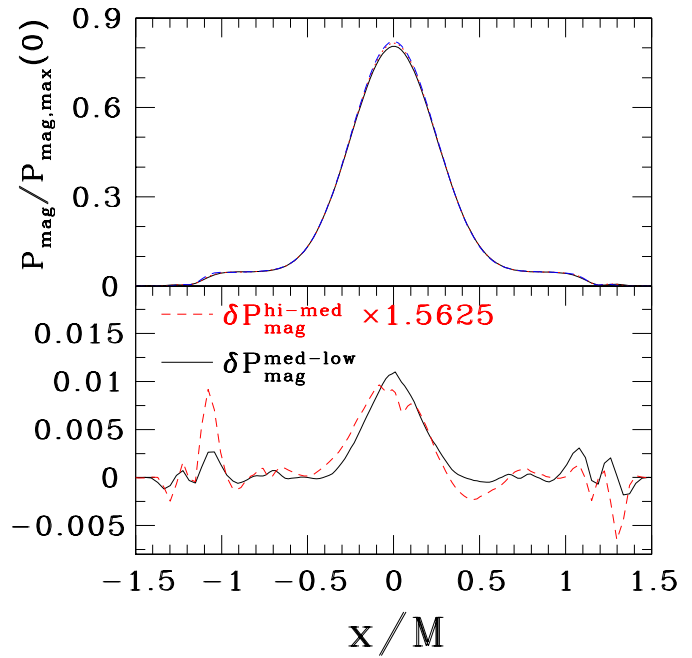}
\caption{Magnetized stellar collapse convergence tests.  Upper graph: Magnetic pressure $P_{\rm mag}=b^2/2$ as a function 
of $x$ along the diagonal line $x=y=z$ at time $t=40.2M$ for the low
(black solid line), medium (red dotted line) and high (blue dashed line) 
resolution runs, normalized by the initial maximum value of $P_{\rm mag}$. 
Lower graph: Pairwise differences of $P_{\rm mag}$ between different 
resolution runs. The difference 
$\delta P_{\rm mag}^{\rm hi-med}=(P^{\rm hi}_{\rm mag}-P^{\rm med}_{\rm mag})
/P_{\rm mag,max}(0)$ is multiplied by 1.5625 to demonstrate deviations from 
second-order convergence. Notice that the results converge slightly higher than 
second order in the high $P_{\rm mag}$ region but less than second order 
in the low $P_{\rm mag}$ region.} 
\label{fig:Pmag_col}
\end{figure}

Since the star is unstable, collapse can be triggered by numerical truncation 
error during the evolution. However, since the truncation error is reduced 
with increasing resolution, 
subsequent evolution of the star will depend sensitively on resolution, which 
is not desirable for a convergence test. We therefore induce the collapse by 
depleting the initial pressure by one percent. We set up a small, poloidal, 
axisymmetric magnetic field by setting the vector potential as follows:
\beqn
  A_x &=& -y A_b \max(P-P_{\rm cut},0) \ , \\ 
  A_y &=& x A_b \max(P-P_{\rm cut},0) \ , \\
  A_z &=& 0 \ ,
\eeqn
where $P_{\rm cut}$ is set to 4\% 
of the initial maximum pressure. The constant parameter $A_b$ determines 
the strength of the magnetic field. We characterize the strength of the 
magnetic field by the ratio of the magnetic energy $\cal M$ to the internal energy 
$E_{\rm int}$. These energies are defined as 
\beqn
 E_{\rm int} = \int \sqrt{-g} (\rho_0 \epsilon) u^0 d^3 x \ , \\ 
  {\cal M} = \int \sqrt{-g} (b^2/2) u^0 d^3 x  \ .
\eeqn
We have chosen a magnetic field strength of ${\cal M}/E_{\rm int}=7.3\times 10^{-3}$, 
which introduces only a small perturbation to the star.

Figure~\ref{fig:alphac} shows the evolution of the central lapse for the three 
resolution runs. As the star is collapsing, the lapse decreases and an apparent 
horizon appears at time $t\sim 70M$. The large energy density and magnetic 
pressure inside the horizon causes the code to crash soon after its formation. 
This difficulty can be overcome by evacuating the hydrodynamic 
matter and magnetic field deep inside the horizon soon after the formation of 
horizon. The evolution then proceeds stably, and the spacetime settles 
to a Kerr black hole after $t \gtrsim 75M$ (see Fig.~\ref{fig:horizon}), 
with virtually no fluid or magnetic fields 
left outside the horizon. The mass and spin of the black hole are computed 
using the isolated and dynamical horizon formalism~\cite{ak04}, with the axial 
Killing vector field computed using the numerical technique described 
in~\cite{dkss03}. We find $M_{\rm BH}\approx M$ and $J_{\rm BH}\approx J$ 
($a_{\rm BH}/M_{\rm BH}=J/M^2=0.34$) for all three resolution runs once all
the matter enters the horizon, where $M$ and $J$ are the initial ADM mass 
and angular momentum of the star, respectively. This is expected 
since the collapse is nearly axisymmetric, and only a negligible amount of mass 
as well as angular momentum is radiated by the gravitational waves. 
(Recall that no angular momentum is radiated in strict axisymmetry.)

Figure~\ref{fig:Pmag_col} shows the profile of magnetic pressure $P_{\rm mag}=b^2/2$ 
along the diagonal line $x=y=z$ at 
$t=40.2M$. We see slightly higher than second-order convergence in the high 
$P_{\rm mag}$ region but lower than second-order in the low 
$P_{\rm mag}$ region. 

In the simulation with the highest resolution, we find that the vector 
potential $A_i$ develop spikes near the second innermost refinement boundary  
during and after the edge of the $A_i=0$ surface passes through that 
refinement boundary. The amplitude of the spikes amplifies with time, 
eventually causing the code to crash. This difficulty can be removed 
by adding a fourth-order Kreiss-Oliger dissipation to $A_i$. The origin of 
the spikes is from prolongation and restriction. As $A_i$ are 
steeply decreasing to zero near the edge, our adopted third-order Lagrangian 
interpolation scheme adds spurious oscillations in $A_i$ near the refinement boundary
after prolongation and restriction. Since the refinement boxes are not moving, 
the oscillation amplitude amplifies each time when prolongation and restriction 
are applied. The same phenomenon could occur for other hydrodynamical variables 
with a steep gradient. However, this effect has a more significant impact on 
the magnetic field, since a slight spatial oscillation in $A_i$ will be amplified 
after taking spatial derivatives. An alternative method to cure this problem
would be to 
use a more sophisticated interpolation scheme such as the ENO
or WENO scheme. We plan to investigate 
these alternative interpolation schemes in the future. 

\section{Conclusion}
\label{sec:summary}

We have developed a new GRMHD code that is capable of evolving MHD
fluids in dynamical spacetimes. We use the BSSN scheme coupled with
the puncture gauge conditions to evolve the metric, and an HRSC scheme
to evolve the MHD and induction equations.

We adopt the formalism
described in~\cite{zbl03} to recast the induction equation 
into an evolution equation for the magnetic vector potential $A_i$ 
[i.e.\ Eq.~(\ref{eq:indAi})]. The variables $A_i$ are 
stored on a staggered grid with respect to the other 
variables. The divergenceless constraint $\ve{\nabla}\cdot \ve{B}=0$ 
is imposed through the vector potential.  This evolution scheme is AMR-compatible, with  
prolongation and restriction applied to the unconstrained 
variables $A_i$ instead of $B^i$, which gives us flexibility in
choosing different interpolation schemes for prolongation/restriction. In simulations with uniform 
grid spacing, our scheme for evolving the magnetic field is numerically 
equivalent to the commonly used constrained-transport scheme based on 
a staggered mesh algorithm~\cite{eh88}.

We have performed several code tests to 
validate our code, including magnetized shocks, nonlinear Alfv\'en waves, 
cylindrical blast explosions, cylindrical rotating disks, magnetized 
Bondi tests, and collapse of magnetized rotating stars. We find good 
agreement between the analytic and numerical solutions, and achieve 
second-order convergence for smooth flows, as expected.

In GRMHD simulations in dynamical spacetimes involving black holes, one 
delicate issue is the handling of the black hole interior. We adopt the 
moving puncture technique in which the black hole spacetime singularity is avoided 
by the puncture gauge conditions. However, a coordinate singularity (puncture) 
remains in the black hole interior, which could cause numerical difficulties 
in MHD simulations. In our tests involving black holes, we find that 
the evolution in the black hole interior is more stable when a more 
diffusive scheme such as the MC reconstruction scheme is used rather 
than the PPM scheme. We plan to investigate the idea of using a less 
diffusive scheme (such as PPM reconstruction coupled with the HLL 
flux) in the black hole exterior and a more diffusive scheme 
(such as MC or minmod reconstruction coupled with the LLF flux) in the 
black hole interior. A similar technique is used 
in some MHD simulations of magnetized accretion disks around 
a black hole~\cite{gammie-priv}. 
We also find that adding Kreiss-Oliger dissipation to MHD variables
in the black hole interior can stabilize the evolution.

In GRMHD simulations using an FMR grid, we find that applying a high order 
interpolation scheme on $A_i$ during prolongation and restriction 
could cause oscillations in $A_i$ near the refinement boundaries. 
The oscillation amplitude can amplify with time. This numerical 
artifact degrades the accuracy of the simulation and could even 
cause the code to crash. The artifact can be removed by adding 
a fourth order Kreiss-Oliger dissipation to $A_i$. A better 
solution is to use a more sophisticated interpolation scheme 
for $A_i$, such as the ENO or WENO scheme. We plan to investigate these 
alternative schemes in the future.

In addition to the treatment of the black hole interior, our MHD code 
has limitations similar to those of other MHD codes in the literature.  In
particular, accurate evolution is difficult when $b^2 \gg \rho_0$.  This 
could potentially cause problems in the low-density regions in some 
applications.  However, our experience and the experience of other numerical 
MHD groups suggests that these difficulties are surmountable.

Having demonstrated the validity of our AMR GRMHD code, we will next apply our code
to study the effects of magnetic fields in the coalescence of binary neutron 
star and black hole-neutron star systems, the collapse of magnetized supermassive 
stars, and the dynamics of magnetized accretion disks around merging binary 
black holes.

\acknowledgments 
This paper was supported in part by NSF
Grants PHY06-50377 and PHY09-63136 as well as NASA
NNX07AG96G and NNX10A1736 to the University of Illinois at
Urbana-Champaign.
Simulations were performed under a TeraGrid Grant TG-MCA99S008
and on the Illinois Numerical Relativity Beowulf Cluster.

\bibliography{paper01}
\end{document}